\documentclass[12pt]{article}

\usepackage[margin=1in]{geometry}
\usepackage{times}

\usepackage{microtype}
\usepackage{graphicx}
\usepackage{booktabs}
\usepackage{amsmath}
\usepackage{amssymb}
\usepackage{amsthm}
\usepackage{mathtools}
\usepackage{natbib}
\usepackage{setspace}
\usepackage{etoolbox}
\AtBeginEnvironment{table}{\setstretch{1.05}\setlength{\tabcolsep}{4pt}\renewcommand{\arraystretch}{1.15}}
\AtBeginEnvironment{thebibliography}{\small\setstretch{1.05}\interlinepenalty=10000}
\usepackage{caption}
\usepackage{subcaption}
\usepackage{array}
\usepackage{float}

\newcommand{\thead}[1]{\shortstack{#1}}
\usepackage{tabularx}
\usepackage{xcolor}
\usepackage{hyperref}

\hypersetup{
    pdftitle={Decision-Relevant Information in Partially Observed Production Networks},
    pdfauthor={},
    colorlinks,
    linkcolor={blue!60!black},
    citecolor={blue!60!black},
    urlcolor={blue!60!black}
}

\newtheorem{lemma}{Lemma}
\newtheorem{proposition}{Proposition}
\newtheorem{corollary}{Corollary}

\newcommand{\Fcal}{\mathcal{F}}
\newcommand{\Ycal}{\mathcal{Y}}
\newcommand{\Ccal}{\mathcal{C}}
\newcommand{\Row}{\operatorname{Row}}

\newcommand{\DomesticSecondMean}{0.360}

\newcommand{\DomesticContraction}{0.5665}
\newcommand{\DomesticSpectralRadius}{0.5239}
\newcommand{\DomesticMaximumRow}{1.1308}
\newcommand{\ImportedInputPercent}{10.38}
\newcommand{\DomesticInputTotal}{407,840,277.7}
\newcommand{\ImportedInputTotal}{47,225,486.3}

\begin{document}
\doublespacing
\clubpenalty=10000
\widowpenalty=10000
\displaywidowpenalty=10000
\setlength{\emergencystretch}{2em}
\renewcommand{\thefootnote}{\fnsymbol{footnote}}

\begin{center}
	{\Large \textbf{Decision-Relevant Information in\\Partially Observed Production Networks}\textsuperscript{*}\par}
	\vspace{1.5em}
	
	\begin{tabular}{>{\centering\arraybackslash}m{0.3\textwidth} >{\centering\arraybackslash}m{0.3\textwidth} >{\centering\arraybackslash}m{0.3\textwidth}}
		\textbf{Shaowen Luo} & \textbf{Kwok Ping Tsang} & \textbf{Zichao Yang} \\
	\end{tabular}
	
	\vspace{0.75em}
	{September 2026\par}
\end{center}

\footnotetext[1]{Luo: Department of Economics, Virginia Tech, \href{mailto:sluo@vt.edu}{sluo@vt.edu}. Tsang: Department of Economics, Virginia Tech, \href{mailto:byront@vt.edu}{byront@vt.edu}. Yang: Wenlan School of Business, Zhongnan University of Economics and Law, \href{mailto:yang\_zichao@outlook.com}{yang\_zichao@outlook.com}.}
\setcounter{footnote}{0}
\renewcommand{\thefootnote}{\arabic{footnote}}

\begingroup
\setstretch{1.3}

\begin{abstract}
A production network can remain largely unidentified even when the economic decision it supports is identified. We characterize sufficient measurements for exposure-based decisions and compute sharp maximum regret over networks consistent with released totals. Using earlier and later vintages of Japan’s interregional input-output accounts, we select measurements from the 1995 table and evaluate the frozen design against the 2005 benchmark. At roughly half the statistics required for full disclosure, the resulting monitoring set loses only 0.07 percentage points of average exposure relative to the benchmark optimum, yet its sharp maximum regret across compatible networks is 4.83 points. In U.S. coal deliveries surrounding a 2005 Wyoming rail disruption, additional shipment measurements identify the optimal set of plants to monitor for inventory risk, even though four monitored plants' exposures to the affected coal supply remain unidentified. The results distinguish good benchmark performance from a decision guarantee and show that decisions can be identified before individual exposures. They suggest evaluating network data by the economic decisions they support.

\end{abstract}
\endgroup



\clearpage

\section{Introduction}
\label{sec:introduction}
Which exposure-based decisions can regional input-output data support when interregional flows are only partially observed? Researchers often complete those flows before measuring exposure to localized shocks. Yet a monitoring set, threshold classification, or exposure ranking may be identified even when many underlying links remain unknown. Conversely, a completed network can yield a precise answer whose validity depends on the completion rule.

We retain every network consistent with the observed releases and basic accounting restrictions. For a given exposure, this set reveals whether the data identify the target and which additional measurements would remove any remaining ambiguity. For a proposed monitoring set, we calculate maximum regret: the largest gap in exposure between the proposed set and the best monitoring set evaluated in the same compatible network. This criterion separates good performance in one benchmark network from a guarantee supported by the released information.

We first apply this approach to Japan's interregional input-output accounts, choosing measurements using the 1995 accounts and evaluating the same design in 2005. The candidate measurements report shocked-origin purchases by destination, buyer group, or buyer industry. We select among them to minimize exposure uncertainty under a fixed statistic budget. At roughly half the statistic count of the full-disclosure template, the selected measurements leave most network links unidentified and exposure intervals that average 2.09 percentage points in width. We find that the resulting monitoring set loses only 0.07 percentage points of average exposure relative to the constructed 2005 benchmark optimum.

Holding these measurements and the monitoring set fixed, however, sharp maximum regret across compatible networks is 4.83 percentage points. The selected set is nearly optimal in the benchmark, but the releases permit substantially larger losses in other networks. Additional measurements therefore improve the decision guarantee without necessarily making it as tight as benchmark performance would suggest.

We next turn to a different setting in which the underlying transactions are directly reported rather than constructed from regional input-output accounts. Following the May 2005 disruption of Wyoming’s Joint Line, we use pre-disruption coal deliveries to study which power plants should be monitored for inventory risk based on their exposure to disrupted coal supply. The application illustrates the converse identification result: a monitoring decision can be identified even when some individual exposures remain uncertain. With additional shipment information, the optimal monitoring set is the same in every compatible network even though four monitored plants’ exposures remain unidentified. A set selected using historical exposure already achieves a nearly optimal monitoring score in the recorded network. Additional information therefore strengthens the decision guarantee, with no demonstrated improvement in predicting subsequent low inventories.

The reported coal transactions complement the constructed Japanese accounts. We return to the latter to examine how sourcing restrictions and sampling uncertainty affect exposure measurement, and whether information sufficient for direct exposure also identifies propagation.

First, requiring buyers within a group to share the same sourcing proportions sharply narrows exposure intervals, but can exclude benchmark exposure. In a controlled sampling experiment, the mean common-sourcing interval shrinks to 0.13 percentage points as sampling noise falls, while its coverage of benchmark exposure declines to 30.0 percent. More precise estimation of the released totals therefore does not correct an inaccurate sourcing restriction.

Second, we ask whether information sufficient for direct exposure also identifies downstream propagation. In a fixed-coefficient domestic cost model, observing every direct response still leaves second-round responses uncertain. The information required therefore depends not only on the network that is observed, but also on the economic object the researcher wants to measure.

Regional input-output analysis has long confronted the fact that interregional flows are less completely observed than regional and industry totals. A large literature therefore estimates or reconciles missing regional transactions \citep{Moses1955,CanningWang2005,BoeroEtAl2018,MillerBlair2009}. \citet{West1981} allocates data collection to improve regional multiplier estimates, while \citet{JiangEtAl2010} evaluates measurement choices using earlier and later regional tables. Our focus differs in evaluating measurements by the economic conclusions they identify and the decision losses they leave, rather than by the accuracy of a completed table.

A related regional literature studies how spatial production linkages transmit localized disruptions and shape regional resilience \citep{TodoNakajimaMatous2015,OosterhavenBouwmeester2016,CarvalhoEtAl2021}. We ask a complementary question: when buyer-specific regional sourcing is only partially observed, which exposure and monitoring conclusions are actually supported by the available information?

Our maximum-regret criterion connects this question to treatment choice under partial identification \citep{Manski2007Regret,Stoye2007Regret}. We use it to evaluate monitoring decisions over all networks consistent with the released information, asking how accounting restrictions and additional measurements tighten the resulting decision guarantee.

Network reconstruction has also studied economic outcomes when links are unobserved, including contagion and systemic risk \citep{MastromatteoEtAl2012,AnandCraigVonPeter2015,IalongoEtAl2022}. We evaluate decisions over the full set of networks consistent with the releases, without assigning probabilities to the unobserved allocations. Related problems of unobserved allocation arise in ecological inference, data combination, and transportation polytopes \citep{DuncanDavis1953,CrossManski2002,Bolker1972}. We also allow released totals to be estimated and study how far sourcing can depart from an assumed pattern before a conclusion fails, connecting the analysis to inference under partial identification and breakdown-frontier methods \citep{ImbensManski2004,ChernozhukovHongTamer2007,Stoye2009,KaidoMolinariStoye2019,MastenPoirier2020}. The distinction between direct and propagated exposure also relates to the role of input linkages in production-network models \citep{AcemogluEtAl2012,CaliendoEtAl2018,BaqaeeFarhi2019}.

In what follows, Section 2 develops the identification and monitoring framework. Sections \ref{sec:japan} through \ref{sec:information} study the Japanese application, first comparing completion methods and then studying which additional measurements reduce exposure uncertainty and monitoring loss. Section \ref{sec:coal} provides a complementary application using reported U.S. coal deliveries. The remaining sections return to the Japanese accounts to examine broader implications: Section \ref{sec:sensitivity} studies sourcing restrictions, Section \ref{sec:use} incorporates sampling uncertainty, and Section \ref{sec:propagation} asks whether information sufficient for direct exposure also identifies downstream propagation.

\section{Exposure, Monitoring, and Released Information}
\label{sec:exposure}

\subsection{Direct exposure and monitoring}

Let $i$ index supplier industries, $r$ supplier regions, $d$ buyer regions, and $j$ buyer industries. The nonnegative flow $F_{irdj}$ is the value of industry $i$ inputs attributed to source category $r$ and purchased by industry $j$ in region $d$. Source categories distinguish producing regions when the accounts separate domestic and imported goods. Section \ref{sec:japan} specifies the convention in the Japanese releases. When four indices are unnecessary, $b=(d,j)$ denotes a buyer and $f$ vectorizes $F$.

Let $D_{dj}=\sum_{i,r}F_{irdj}>0$ denote the buyer's total intermediate purchases, and write $D_b=D_{dj}$. The supplier weights $q_{ir}$ specify the disturbance. A binary regional shock assigns one to affected suppliers and zero elsewhere. Direct exposure is
\begin{equation}
E_{dj}(f,q)
=
\frac{\sum_{i,r}q_{ir}F_{irdj}}{D_{dj}}.
\label{eq:direct_exposure}
\end{equation}
We observe $D_{dj}$ in the Japanese application. For a fixed buyer, exposure is therefore a linear target $c'f$, where $c$ collects the supplier weights normalized by that buyer's purchases.

A decision maker has capacity $K$ and assigns the same monitoring cost to each buyer. Holding $q$ fixed, the monitoring score is $V(S,f)=\sum_{b\in S}E_b(f,q)$, and the best set $S_K(f)$ contains the $K$ largest exposures. This objective prioritizes vulnerability measured by affected input shares. A spending objective would instead weight exposures by purchases and generally select different buyers. We use stable region-sector order to resolve ties.

For a proposed set $S$, the loss in average monitored exposure is
\begin{equation}
\ell(S,f)=\frac{100}{K}\left\{\max_{T:|T|=K}V(T,f)-V(S,f)\right\}.
\label{eq:monitoring_loss}
\end{equation}
The factor $100/K$ expresses loss in percentage points of average exposure across the monitored buyers. The relative benchmark score is $V(S,f^{\mathrm{bench}})/V(S_K(f^{\mathrm{bench}}),f^{\mathrm{bench}})$. Overlap measures which buyers are selected, while score loss measures the exposure forgone.

\subsection{Released totals and completed networks}

An exact release is a linear map
\begin{equation}
y=Af.
\label{eq:release_operator}
\end{equation}
Each row of $A$ specifies an observed sum of transactions. These sums may be national industry flows, supplier-region totals, buyer-region-industry totals, regional bilateral totals, or selected cross-tabs. Together with nonnegativity and valid support restrictions, the release defines
\begin{equation}
\Fcal(y)=\{f\geq0:Af=y\}.
\label{eq:baseline_set}
\end{equation}
For an exposure coefficient $c$, the identified set is
\begin{equation}
[L_c(y),U_c(y)]
=
\{c'f:f\in\Fcal(y)\}.
\label{eq:identified_set}
\end{equation}
We obtain the endpoints by minimizing and maximizing exposure over the feasible networks, using linear programs. A threshold $c'f\geq\kappa$ follows from the release when $L_c(y)\geq\kappa$. One buyer ranks above another when the sharp lower bound on their exposure difference is positive.

A completion rule is a map $\widehat f^{\,m}=T_m(y)$ that selects one table from the feasible set, or a nearby table when its objective is approximate. Methods sharing the same sourcing restriction may agree even when the release leaves exposure unidentified.

If supplier weights are constant across source categories, buyer margins identify the corresponding industry input share. When shock weights differ across origins, exposure also depends on which origins supply each buyer. For a proposed monitoring set, maximum regret is the largest monitoring loss across networks consistent with the release:
\begin{equation}
R(S;y)=\max_{f\in\Fcal(y)}\ell(S,f).
\label{eq:worst_monitoring_loss}
\end{equation}
Both the proposed set and its comparator use the same feasible network. Here $R(S;y)$ evaluates a specified monitoring set. Minimax regret is obtained by minimizing it over sets of size $K$. Our main comparisons evaluate sets selected by completion rules. Appendix \ref{app:monitoring_choice} examines whether alternative monitoring sets improve the guarantee at one fixed release. For a fixed set $S$, additional exact releases can only reduce $R(S;y)$.

\section{Japanese Data and Monitoring Design}
\label{sec:japan}

We begin with Japan's interregional input-output accounts, which provide detailed regional-sector transactions that can serve as a benchmark while allowing us to study what remains identifiable when buyer-specific regional flows are withheld. The data are especially useful for localized supply disruptions, where exposure depends on which regions supply which downstream buyers. We use the accounts to construct the benchmark exposures, define progressively richer information releases, and evaluate the monitoring decisions supported by each release.

\subsection{Benchmark table and exposure targets}

The benchmark is the official 2005 METI nine-region, 29-sector interregional input-output table \citep{METI2005}. METI constructs the detailed regional allocation using common sourcing coefficients across purchasing industries, then aggregates and balances the accounts \citep[p.~131]{METIConstruction2010}. Published buyer sectors combine detailed inputs in different proportions, and balancing can further alter their recorded sourcing shares. The construction assumption is therefore not identical to common sourcing within our aggregated buyer groups. Detailed common sourcing may favor related completion methods, although the direction and magnitude of its effect on our comparisons have not been established. The coal application in Section \ref{sec:coal} provides complementary evidence from reported deliveries, subject to its retrospective sample and the assumptions underlying its monitoring score.

\begin{samepage}
The source uses competitive foreign-import accounting. Within-region cells combine locally produced inputs and imported goods used there. Cross-region cells contain domestic interregional purchases \citep[pp.~121--122]{METIConstruction2010}. Write the published array as $F^{\mathrm{tab}}_{irdj}=F^D_{irdj}+\mathbf1\{r=d\}M_{idj}$, where $M$ denotes imported input use. Direct targets are outside the shocked region, so their numerators use domestic cross-region flows and their denominators include all intermediate inputs. We calculate releases from $F^{\mathrm{tab}}$, so the source-category totals include imports attributed to within-region use. Section \ref{sec:propagation} constructs $F^D$ separately for domestic cost propagation.\par
\end{samepage}

Transaction values are measured in million yen. We set nine negative balancing cells, totaling 537 million yen, to zero and calculate releases from that nonnegative array. Appendix \ref{app:data} gives the source mappings, accounting reconciliation, and the common 26-sector panel used for historical comparisons.

For a shock to supplier family $\mathcal S$ in origin $r$, the benchmark target outside that origin is
\begin{equation}
E_{dj}^{\mathrm{bench}}(r,\mathcal S)
=
\frac{\sum_{i\in\mathcal S}F_{irdj}}{\sum_{i,r'}F_{ir'dj}}.
\label{eq:japan_target}
\end{equation}
We cross all nine origin regions with three supplier families, defined by sector roles before comparing completions. All manufacturing contains sector codes 30 through 180, materials contains codes 70 through 130, and machinery contains codes 140 through 170.

We focus on a hypothetical disruption to Tohoku manufacturing. The choice of region and sector is motivated by the Great East Japan Earthquake, but the exercise uses the 2005 production network as a benchmark and does not seek to estimate the earthquake's realized effects.

We focus on a hypothetical disruption to Tohoku manufacturing and its downstream buyers, motivated by evidence from the Great East Japan Earthquake on supply-chain propagation \citep{BoehmEtAl2019,CarvalhoEtAl2021}. The exercise uses the 2005 production network as a benchmark and does not seek to estimate the earthquake's realized effects.

\subsection{Aggregate releases and completion methods}

For each supplier sector, the baseline release retains national supplier-origin totals $o_{ir}=\sum_{d,j}F_{irdj}$ and purchases by each buyer $c_{idj}=\sum_rF_{irdj}$. Origin totals by destination are initially withheld. Richer releases add origin--destination totals or origin--destination--buyer-group totals. For a single shocked origin, the direct bounds use its totals versus the rest of the origins. Point completions use the full origin cross-tabs.

We divide buyers into three groups. Goods production contains agriculture through manufacturing. Market and network services contain public utilities, commerce, finance, real estate, transport, information, and business services. Local and public services contain construction, public administration, education and research, medical and social services, personal services, and Others.

We compare four point completions.
\begin{enumerate}
    \item \textit{Proportional margins} allocates each supplier's regional total across buyers in proportion to buyer margins.
    \item \textit{Gravity-RAS margins} starts from a distance prior and balances to the same origin and buyer margins. Gravity coefficients are calibrated on nonmanufacturing suppliers.
    \item \textit{Collapsed bilateral} observes origin-destination totals and uses a common sourcing share within each destination.
    \item \textit{Buyer-group bilateral} observes origin-destination-buyer-group totals and uses a common sourcing share within each group.
\end{enumerate}
Manufacturing buyer-specific flows are withheld from completion and used for evaluation. Appendix \ref{app:computation} specifies the Gravity-RAS training fields and fitting criterion.

\subsection{Monitoring rule and losses}

The decision maker can monitor twenty buyer region-industries. This capacity is close to one-tenth of the 232 outside-origin targets and is set before evaluating the methods. Appendix Table \ref{tab:capacity_sensitivity} reports capacities of ten and thirty.

For each method we compare pairwise rankings, membership of the benchmark top twenty, and the selected set's exposure score relative to the benchmark optimum. We also compare classifications at the 2 and 3 percent exposure thresholds. Pairwise comparisons exclude exact benchmark ties.

\section{When Does Completion Change the Decision?}
\label{sec:completion}

Completion methods differ mainly in how they split a shocked region's sales among buyers in the same group. We summarize this choice by a sourcing restriction: each buyer's purchase share from the shocked region lies within a deviation $\delta$ of its group's average share, so that $\delta=0$ imposes a common share and larger $\delta$ allows more heterogeneity. The breakdown point $\delta^*$ of a conclusion is the largest $\delta$ for which the conclusion holds in every compatible network. Section \ref{sec:sensitivity} formalizes both the restriction and the breakdown point. This section asks when completion changes the monitoring decision, first for individual Tohoku conclusions and then across all localized designs.

\subsection{Classifications under the Tohoku manufacturing shock}

For a Tohoku shock, some buyer conclusions follow from the group release alone while others are fragile to small sourcing deviations. Table \ref{tab:motivating_conclusions} reports four such cases, each with its completed and benchmark exposure and its breakdown point $\delta^*$.

\begin{table}[H]
\centering
\caption{Completed-network conclusions in the Tohoku case}
\label{tab:motivating_conclusions}
\small\setstretch{1.05}
\begin{tabular}{@{}>{\raggedright\arraybackslash}p{5.2cm}crrr@{}}
\toprule
Buyer & Conclusion & \thead{Completed\\exposure (\%)} & \thead{Benchmark\\exposure (\%)} & \thead{Breakdown\\$\delta^*$} \\
\midrule
Kanto transportation equipment & $>2\%$ & 3.37 & 3.57 & 1.0000 \\
Hokkaido electrical machinery & $>3\%$ & 4.71 & 4.79 & 0.1328 \\
Hokkaido timber and furniture & $<2\%$ & 1.83 & 3.40 & 0.0041 \\
Hokkaido precision instruments & $>3\%$ & 3.80 & 2.97 & 0.0147 \\
\bottomrule
\end{tabular}
\begin{minipage}{0.96\textwidth}
\vspace{0.4em}
\footnotesize
\textit{Notes}: Completed exposure uses observed Tohoku-by-buyer-group flows and a common Tohoku sourcing share within each group. The benchmark is the held-out 2005 buyer-specific table. The breakdown point $\delta^*$ is the largest deviation allowed by \eqref{eq:binary_restriction} under which the stated conclusion holds for every compatible allocation. A value of one means the group release alone identifies the conclusion.
\end{minipage}
\end{table}

Kanto transportation-equipment exposure above 2 percent follows from the group release. Hokkaido electrical-machinery exposure above 3 percent requires the sourcing restriction, but survives a deviation of 0.133. The timber and furniture and precision-instruments conclusions are incorrect in the held-out table and fail at deviations of 0.004 and 0.015.

\subsection{Performance across regions and supplier families}

We now extend the comparison beyond the Tohoku manufacturing case to all 27 combinations of nine origin regions and three supplier families. Across these localized shock designs, the buyer-group method performs best on the average measures we report (Table 2, Panel A). It reduces mean pairwise errors to 10.3 percent and recovers 79.8 percent of the benchmark monitoring set, although overlap falls to 60 percent in its weakest design. Its selected set's total benchmark exposure averages 92.5 percent of the benchmark top-twenty total. It makes 235 classification errors among 6,264 target-design pairs at the 2 percent threshold.

\begin{table}[!htbp]
\centering
\caption{Completion performance across localized supply shocks}
\label{tab:systematic_performance}
\small
\textit{Panel A. All localized designs}\par\vspace{0.3em}
\begin{tabular}{@{}lrrrrr@{}}
\toprule
Method & \thead{Pairwise\\errors (\%)} & \thead{Mean\\overlap (\%)} & \thead{Minimum\\overlap (\%)} & \thead{Relative\\score (\%)} & \thead{Errors at\\2\%} \\
\midrule
Proportional margins & 21.2 & 50.0 & 25.0 & 69.0 & 914 \\
Gravity-RAS margins & 21.4 & 49.8 & 25.0 & 72.4 & 735 \\
Collapsed bilateral & 12.3 & 70.2 & 40.0 & 87.6 & 271 \\
Buyer-group bilateral & 10.3 & 79.8 & 60.0 & 92.5 & 235 \\
\bottomrule
\end{tabular}
\par
\vspace{0.9em}
\textit{Panel B. Tohoku manufacturing monitoring decision}\par\vspace{0.3em}
\begin{tabular}{@{}lrr@{}}
\toprule
Method & \thead{Benchmark buyers\\selected (of 20)} & \thead{Relative\\score (\%)} \\
\midrule
Proportional margins & 7 & 61.6 \\
Gravity-RAS margins & 6 & 47.8 \\
Collapsed bilateral & 14 & 92.5 \\
Buyer-group bilateral & 15 & 93.6 \\
\bottomrule
\end{tabular}
\begin{minipage}{0.97\textwidth}
\vspace{0.5em}
\footnotesize
\textit{Notes}: Panel A averages over nine origin regions and three supplier families. Minimum overlap is the lowest across these designs. Each design contains 232 positive-purchase buyers outside the shocked origin. Pairwise errors exclude benchmark ties. Relative score is total benchmark exposure in all twenty selected buyers divided by the benchmark top-twenty total, including exposure of selected buyers outside the benchmark top twenty. The 2 percent column counts errors across 6,264 target-design pairs per method. Panel B reports how many benchmark buyers are selected in the Tohoku manufacturing design and the resulting relative score.
\end{minipage}
\end{table}

Proportional margins and gravity-RAS both recover about half of the benchmark top twenty. Gravity selects buyers with greater total exposure and makes fewer threshold errors, yet has slightly more pairwise reversals. The buyer-group method's mean overlap ranges from 73.3 percent for materials to 89.4 percent for machinery (Appendix Table \ref{tab:family_results}). Appendix Figure \ref{fig:completion_conditions} relates its errors to sourcing heterogeneity and shock size.

For Tohoku manufacturing (Panel B), the buyer-group completion selects fifteen of the twenty benchmark buyers and achieves 93.6 percent of the optimum score.

The comparisons show that completion can change both the selected buyers and threshold classifications. Small sourcing deviations are enough to overturn some conclusions, which raises the question of what additional measurements would establish them.

\section{Additional Measurements and Monitoring Loss}
\label{sec:information}

Which additional measurements resolve exposure uncertainty, and how should an agency choose them under a budget? We begin with a threshold conclusion identified by one cross-tab cell, then characterize the information required for exposure targets and evaluate measurements selected using earlier accounts.

\subsection{Identifying a threshold with one cross-tab}

The Kanto transportation-equipment conclusion in Table \ref{tab:motivating_conclusions} requires only one cross-tab cell beyond national supplier and buyer margins. In the transportation-equipment supplier block, the released Tohoku flow to Kanto goods-producing buyers is 376,965 million yen. The target buyer purchases 9,347,221 million yen in this block, while the goods-group total is 9,359,510 million yen. The Fr\'{e}chet lower bound, the smallest flow consistent with the released totals, forces at least
\begin{equation}
376{,}965+9{,}347{,}221-9{,}359{,}510
=364{,}676
\label{eq:certificate_flow}
\end{equation}
million yen into the target buyer. Dividing by total intermediate purchases gives
\begin{equation}
\frac{364{,}676}{16{,}097{,}139}
=0.02265>0.02.
\label{eq:certificate_exposure}
\end{equation}
In the common 26-sector panel, each candidate cell reports one manufacturing supplier's Tohoku sales to the destination and group containing the target buyer. In 142 cases, releasing selected cells determines which side of the threshold the exposure falls on, mostly below it, and 36 of these need only one cell. Kanto transportation equipment is again the only buyer whose exposure is shown to exceed 2 percent. Appendix \ref{app:measurement_certificates} derives the minimum cell count and gives dual certificates.

\subsection{What existing releases identify}

When does a release determine exposure itself, beyond identifying a threshold conclusion? The classical estimable-function condition asks whether the target can be formed from the released sums \citep{Searle1965}.

\begin{samepage}
\begin{lemma}[Linear estimability on realized support]
\label{lem:rowspace}
Suppose $\Fcal(y)$ is nonempty, and let $\mathcal E$ be its effective support: the coordinates that are positive at a relative-interior point of its minimal face. Then $c'f$ is constant on $\Fcal(y)$ if and only if $c_{\mathcal E}\in\Row(A_{\mathcal E})$.
\end{lemma}
\end{samepage}

Effective support includes every transaction that is positive in some feasible table. A zero in the published benchmark or a completion does not by itself remove a transaction from this support. Accounting, technology, or institutional restrictions must exclude it.

To see what margins identify, suppress supplier industry and combine destination and buyer industry into one index $k$. Suppose the release contains origin margins $o_r=\sum_kF_{rk}$ and buyer margins $d_k=\sum_rF_{rk}$.

\begin{lemma}[Margin-invariant coefficients]
\label{lem:additive}
On an effective bipartite support $\mathcal E$, the functional $\sum_{(r,k)\in\mathcal E}C_{rk}F_{rk}$ is invariant to every margin-preserving reallocation if and only if
\begin{equation}
C_{rk}=\alpha_r+\beta_k
\label{eq:additive_coefficients}
\end{equation}
on each connected component of $\mathcal E$.
\end{lemma}

Let $a_k$ be the weight assigned to buyer $k$ in the target, including any normalization by purchases. For a rank-one coefficient $C_{rk}=q_ra_k$ on complete support, the identifying condition fails when both the supplier shock and buyer weights vary. The relevant cycle contrast is
\begin{equation}
C_{rk}-C_{rk'}-C_{r'k}+C_{r'k'}
=
(q_r-q_{r'})(a_k-a_{k'}).
\label{eq:rank_one_cycle}
\end{equation}
Margins identify exposure when the shock is uniform across origins or buyer weights are equal. With both weights varying, a swap of size $\epsilon$ around this cycle changes the target by $\epsilon(q_r-q_{r'})(a_k-a_{k'})$. The size of a feasible swap depends on the purchases that other buyers can absorb.

Shocked-origin cross-tabs identify exposure aggregated by destination or buyer group, but can leave individual exposures uncertain because purchases can be reallocated within a group. To assess a pairwise ranking, we apply the same argument to the difference between the two buyers' exposure coefficients. Each additional cross-tab rules out reallocations that would change its reported total. The remaining reallocations lie in the null space of the augmented release operator. Appendix \ref{app:network_geometry} shows why network density and cycle count alone do not order exposure uncertainty.

\subsection{Information needed for an exposure family}

One tailored statistic identifies a fixed exposure. How many are needed when the shock or buyer weights may vary?

Fix an effective support and restrict the baseline operator $A$ and all coefficients to it. We suppress the support subscripts. Let $\Ccal=\operatorname{span}\{c_1,\ldots,c_P\}$ contain a prespecified family of exposure coefficients.

\begin{proposition}[Additional statistics for a target family]
\label{prop:target_family_rank}
On a fixed effective support, the minimum number of independent additional release rows needed to identify every coefficient in $\Ccal$ is
\begin{equation}
r^*
=
\dim(\Ccal)-\dim(\Ccal\cap\Row(A)).
\label{eq:target_family_rank}
\end{equation}
Additional rows attain this bound when their span together with $\Row(A)$ contains $\Ccal$. Fewer than $r^*$ rows cannot identify the full family.
\end{proposition}

Suppose effective support is complete and the baseline contains only row and column margins. Group origins into $G$ classes with common shock weights and buyers into $H$ classes with common target weights. The margins leave $(G-1)(H-1)$ independent flows between these classes unobserved. Releasing them identifies every exposure whose coefficients are constant within each class pair, while leaving allocations within those pairs unknown. Restricted support or additional baseline releases can reduce the count. Equation \eqref{eq:target_family_rank} gives the corresponding support-restricted rank.

\subsection{Choosing measurements using earlier accounts}

Proposition \ref{prop:target_family_rank} permits arbitrary additional linear measurements. An agency restricted to particular fields faces a different problem. Let $\mathcal{G}=\{g_1,\ldots,g_L\}$ be candidate release rows or cross-tab blocks, and let $G_{\mathcal A}$ stack those chosen in $\mathcal A\subseteq\{1,\ldots,L\}$. Given the population releases $(Af_0,G_{\mathcal A}f_0)$, let $\operatorname{wid}_t(\mathcal A)$ be the sharp width of target $t$. With target weights $\omega_t$ and release costs $k_\ell$, the design solves
\begin{equation}
\mathcal A_B^*
\in
\arg\min_{\mathcal A:\,\sum_{\ell\in\mathcal A}k_\ell\leq B}
\sum_{t=1}^P\omega_t\operatorname{wid}_t(\mathcal A).
\label{eq:release_design_problem}
\end{equation}
Alternative objectives count unresolved threshold classifications or rankings. A menu may require redundant rows or exclude those attaining $r^*$, so its minimum cost depends on the available fields and their prices. When cross-tabs overlap, we count only restrictions that are not implied by earlier releases: the increase in rank when their rows are added to the release operator.

Evaluating an existing release uses its observed values, whereas choosing additional fields requires a criterion that can be applied before their values are known. We use the 1995 accounts to select fields and the 2005 accounts to evaluate them.

The common 26-sector panel has 208 positive-purchase buyers outside Tohoku and 14 manufacturing suppliers. Each supplier has four nested release levels: no additional information, or Tohoku purchases by destination, destination and buyer group, or destination and individual buyer industry. Their fixed template costs are 0, 8, 26, and 233 statistics beyond national supplier-origin and buyer margins, giving 3,262 for the full menu. The counts depend on the sector and region classifications and are fixed before allowing for flows that the accounts force to zero. They do not measure confidentiality costs. Appendix \ref{app:network_measurement} reports effective ranks and raw cell counts.

\begin{figure}[H]
\centering
\includegraphics[width=0.98\textwidth]{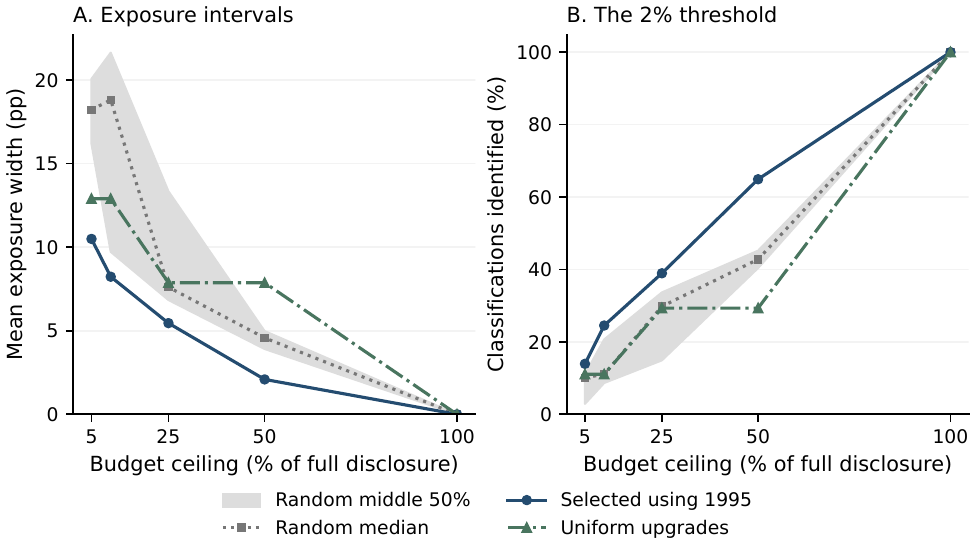}
\caption{Exposure uncertainty under measurements selected in 1995}
\label{fig:network_measurement}
\begin{minipage}{0.95\textwidth}
\vspace{0.3em}
\footnotesize\setstretch{1.1}
\textit{Notes}: Fields are selected using 1995 and evaluated on 208 outside-Tohoku buyers in 2005. Budget ceilings are fractions of 3,262 template statistics, and actual use may be lower. Uniform upgrades give every supplier the same release level. Random comparisons use 200 feasible allocations per budget. The shaded area is their interquartile range, not a confidence band. Panel B counts sharp intervals entirely on one side of the 2 percent threshold. The width-per-cost rule selects the same fields as the optimum at four of the five budgets (Table \ref{tab:network_measurement_full}).
\end{minipage}
\end{figure}

We minimize unweighted mean exposure width in 1995. Because supplier blocks are separable, their width contributions are additive and the menu can be optimized by dynamic programming. Figure \ref{fig:network_measurement} evaluates the selected releases in 2005 at budget ceilings of 5, 10, 25, 50, and 100 percent of the full cost. Panel A measures remaining exposure uncertainty, while Panel B reports the share of 2 percent classifications identified. Appendix \ref{app:measurement_allocation} gives the recursion, alternative allocation rules, and the experimental 2000 comparison.

\begin{samepage}
At half the budget, the selected release uses 1,606 statistics and leaves mean width 2.09 percentage points, compared with 4.57 for the random median and 7.87 for uniform upgrades. It identifies 64.9 percent of the 2 percent classifications and 72.1 percent of the 3 percent classifications. Uniform upgrades use only 364 statistics at this budget, so part of their disadvantage is unused capacity.
\end{samepage}

\begin{samepage}
Historical selection does not always minimize later width. At the 10 percent budget, 2 of 200 random allocations have smaller 2005 mean width. The best gives 8.184 points, compared with 8.230 for the historical optimum. At 25 percent, 1 of 200 does so, giving 5.047 rather than 5.454 points.

\end{samepage}

The half-budget release leaves 46,858 of the original 48,464 independent directions in which network flows can vary. Both counts are calculated before allowing for flows that the accounts force to zero. However, even its narrower intervals leave top-twenty membership unresolved. Across the historical measurement comparisons, each chosen buyer under a partial-budget release can be ranked below at least twenty competitors in a feasible network. Different buyers may require different networks.

\subsection{Monitoring loss under the selected releases}

Unresolved membership can coexist with a small loss when buyers near the cutoff have similar exposures. For each menu, the agency selects twenty buyers by allocating the released 2005 flows proportionally within supplier partitions. Table \ref{tab:monitoring_loss} compares benchmark loss with maximum regret in \eqref{eq:worst_monitoring_loss}. Appendix \ref{app:monitoring_calculation} gives the optimization and attaining allocations.

\begin{table}[!htbp]
\centering\small
\caption{Monitoring overlap, benchmark loss, and maximum regret}
\label{tab:monitoring_loss}
\begin{tabular}{@{}rrrrr@{}}
\toprule
\thead{Budget ceiling\\(\%)} & \thead{Additional\\statistics} & \thead{Overlap\\(\%)} & \thead{Benchmark loss\\(pp)} & \thead{Maximum regret\\(pp)} \\
\midrule
0 & 0 & 55 & 0.831 & 67.618 \\
5 & 158 & 75 & 0.181 & 27.553 \\
10 & 310 & 70 & 0.287 & 21.582 \\
25 & 778 & 80 & 0.147 & 17.606 \\
50 & 1,606 & 90 & 0.070 & 4.834 \\
100 & 3,262 & 100 & 0.000 & 0.000 \\
\bottomrule
\end{tabular}
\begin{minipage}{0.97\textwidth}
\vspace{0.4em}\footnotesize
\textit{Notes}: Common 26-sector panel, 208 outside-Tohoku targets, and capacity twenty. Overlap is the percentage of benchmark top-twenty buyers selected. Both loss columns use percentage points of mean exposure across monitored buyers. Budget ceilings are relative to the full menu's 3,262 template statistics. Fields are selected using 1995 and evaluated in 2005 under accounting restrictions only. Maximum regret uses only the released information and specified shock, while benchmark loss also uses withheld exposures. The maximum is taken across compatible networks, and every unrounded bound is attained to within $10^{-5}$ points.
\end{minipage}
\end{table}

At half the budget, the rule selects eighteen of the twenty benchmark buyers and loses only 0.07 percentage points of average exposure. However, the same releases permit a loss of 4.834 points, compared with 67.618 under national and buyer margins alone. The releases minimize historical interval width, not maximum regret. Moreover, they change the selected set, so benchmark loss need not fall monotonically, as the 5 and 10 percent budgets illustrate.

Holding the half-budget release fixed, we also choose monitoring sets to reduce maximum regret (Appendix \ref{app:monitoring_choice}). The best set found reduces maximum regret from 4.834 to 3.285 points, although benchmark loss rises from 0.070 to 0.783 points. The sets share 10 buyers. The bounds on minimax regret are [2.615, 3.286] points, so the search does not establish global optimality. The comparison optimizes monitoring conditional on this release and does not establish an optimal disclosure policy.

\section{Coal Deliveries and Inventory Monitoring}
\label{sec:coal}

Can a release establish the optimal monitoring set while leaving its buyers' exposures uncertain? We use reported coal deliveries before the May 2005 disruption of Wyoming's Joint Line to examine this question. Derailments and subsequent repairs reduced shipments through much of that year \citep[pp.~82--83]{NRC2007Coal}. Campbell County, Wyoming, provides a geographic proxy for affected supply. County origin does not establish use of a particular rail segment.

\subsection{Data and disclosure design}

The public FERC-423 and EIA-423 records report deliveries by origin and plant. EIA-906/920 supplies inventories and consumption.\footnote{U.S. Energy Information Administration, ``Historic Form EIA-423 \& FERC-423 Detailed Data,'' \url{https://www.eia.gov/electricity/data/eia423/}, and ``Form EIA-923 Detailed Data with Previous Form Data (EIA-906/920),'' \url{https://www.eia.gov/electricity/data/eia923/}. We use the 2004 and 2005 archived files, accessed September 2026.} We aggregate January--April 2005 receipts into a network of 338 plants and 180 positive origin categories, measured in coal energy. The fuel codes retain synthetic coal and exclude petroleum coke, wood, and other refuse. Foreign and unknown origins remain separate, and months without receipt entries contribute zero to reported totals.

We withhold fields from the final archived delivery records to evaluate alternative disclosure policies. The sample is selected retrospectively to include plants with the required generation and stock observations. The accompanying supplement, \textit{Coal Deliveries: Construction and Supplementary Results}, describes the sample restrictions and data construction.

Let $x_j$ denote Campbell receipts, $D_j$ total coal receipts, and $W_j$ Wyoming receipts at plant $j$. Exposure is $e_j=x_j/D_j$. The baseline retains every origin total, every plant's total receipts, and $W_j$. Additional releases disclose Campbell totals by destination region, state, state and sector, or plant. Within each released group $g$, feasible allocations satisfy $0\leq x_j\leq W_j$ and the observed group total. Every Campbell allocation satisfying these constraints can be completed by allocating the other Wyoming and non-Wyoming flows while preserving the baseline totals. The reduced problem therefore gives exact exposure bounds for the full network. We keep inventories separate because the receipts and consumption schedules do not reconcile exactly.

We choose releases to minimize mean exposure width in annual 2004 data, with geography and sector labels fixed in that year. Current baseline information determines which rows are redundant, without using withheld county allocations or subsequent outcomes. In particular, 206 plants have zero Wyoming purchases, so their Campbell exposures are already known. Full exposure disclosure requires 131 additional independent statistics for the remaining 132 plants.\footnote{The Wyoming baseline itself adds 337 independent statistics to origin and plant-total margins. The reported information savings are conditional on retaining Wyoming purchases. They do not establish savings relative to direct Campbell disclosure from basic margins. The counts measure neither reporting nor confidentiality costs.}

\subsection{Monitoring and decision guarantees}

Inventories distinguish plants that can absorb a delivery interruption from those closer to exhausting their stocks. We select twenty plants with the largest scores
\begin{equation}
s_j=H e_j-d_j,\qquad H=30,
\label{eq:coal_score}
\end{equation}
where $d_j$ is end-of-January inventory divided by average daily coal consumption in June--August 2004. The term $H e_j$ expresses the Campbell share of a 30-day consumption requirement in days. The score is the negative of remaining inventory coverage after this hypothetical disruption, assuming fixed consumption, no replacement of Campbell deliveries, and a comparable coal mix. Higher scores indicate smaller inventory buffers. We restrict attention to the 168 plants with at least 30 days of January coverage, of which 48 fall below that level during June--August. The horizon and threshold are exploratory choices, not regulatory standards. The score does not predict electricity interruption.

\begin{samepage}
The completion allocates each group's Campbell total in proportion to its plants' Wyoming purchases. Loss is the gap in mean score between the selected set and the best twenty-plant set in the same network. Table \ref{tab:coal_monitoring} compares this rule with historical exposure ranking. Appendix \ref{app:monitoring_calculation} gives the calculation with the inventory term.\par
\end{samepage}

\begin{table}[H]
\centering\small
\caption{Information and monitoring guarantees in coal deliveries}
\label{tab:coal_monitoring}
\begin{tabular}{@{}rcrrr@{}}
\toprule
\thead{Additional\\statistics} & \thead{Unidentified exposures\\All / monitored} & \thead{Recorded loss\\(days)} & \thead{Maximum regret\\(days)} & \thead{Low-stock\\cases} \\
\midrule
\multicolumn{5}{@{}l}{\textit{Panel A. Proportional completion and January inventories}} \\
\addlinespace[0.25em]
0 & 132 / 20 & 0.054 & 14.654 & 13 \\
27 & 83 / 11 & 0.000 & 3.387 & 14 \\
62 & 44 / 5 & 0.098 & 2.408 & 14 \\
102 & 24 / 4 & 0.000 & 0.000 & 14 \\
131 (full) & 0 / 0 & 0.000 & 0.000 & 14 \\
\addlinespace[0.45em]
\multicolumn{5}{@{}l}{\textit{Panel B. 2004 exposure and January inventories}} \\
\addlinespace[0.25em]
0 & 132 / 20 & 0.058 & 14.588 & 14 \\
\bottomrule
\end{tabular}
\begin{minipage}{0.97\textwidth}
\vspace{0.4em}\footnotesize\setstretch{1.1}
\textit{Notes}: All comparisons retain origin totals, plant-total receipts, and plant Wyoming purchases. Unidentified exposures are counted among all 338 plants and, after the slash, among the twenty selected plants. Panel A uses the historical measurement menu. The 62-statistic release has a budget ceiling of 65. Panel B adds no county releases and uses 2004 exposure to rank plants. Loss is in days of mean score across selected plants. Low-stock cases are selected plants among the 48 that fall below 30 days of coverage in June--August. Full disclosure identifies Campbell exposures, not the entire network.
\end{minipage}
\end{table}

\begin{samepage}
With 27 additional statistics, the selected set matches the fully observed set, but another feasible network permits a mean score loss of 3.387 days. At 102 statistics, maximum regret is zero even though 24 exposures remain uncertain, including four among the monitored plants. Gibbons Creek's Campbell share, for example, ranges from 90.48 to 100 percent without changing the optimality of the selected set. This is the first release along the historical menu to guarantee zero loss. It uses 77.9 percent of the statistics needed to identify all remaining exposures, although another disclosure policy could require fewer.
\end{samepage}

Historical ranking in Panel B loses only 0.058 days in the recorded network and finds the same 14 subsequent low-stock cases as full current disclosure. Additional reports chiefly strengthen the decision guarantee, with no demonstrated forecasting gain. A simpler rule that greedily selects individual plant reports using historical width reductions also establishes optimality at 102 statistics. The supplement compares further reporting rules.

\begin{samepage}
The 102-statistic release also establishes optimality when the set is reselected using disruption horizons of 10 and 90 days. Allowing each inventory measure to differ by one day raises maximum regret to 0.064 days, the same bound obtained with full Campbell disclosure.\par
\end{samepage}

The coal release identifies the optimal monitoring set while leaving some exposures uncertain. In Japan, even the historically selected measurements leave substantial maximum regret. We next ask how sourcing restrictions change the conclusions supported by those releases.

\section{Sensitivity to Sourcing Restrictions}
\label{sec:sensitivity}

How much precision does common sourcing supply beyond the released totals? We return to the Japanese 29-sector table and relax common sourcing toward accounting information alone. We then use the 26-sector panel to calibrate an allowed deviation in 1995 and evaluate it in 2005.

\subsection{Sourcing restrictions and exposure bounds}

Fix supplier industry $i$, destination $d$, and buyer group $g$. Let $c_j$ be buyer $j$'s purchases of supplier $i$, and let $G_g=\sum_{j\in g}c_j$. Let $g_T$ denote the group's observed purchases from the shocked origin. The pooled sourcing share is
\begin{equation}
\bar p_g=\frac{g_T}{G_g}.
\label{eq:pooled_share}
\end{equation}
For buyer $j\in g$, let $x_j$ be its latent purchases from the shocked origin and $p_j=x_j/c_j$. We impose
\begin{equation}
|p_j-\bar p_g|\leq\delta,
\qquad 0\leq\delta\leq1.
\label{eq:binary_restriction}
\end{equation}
The restriction concerns only the shocked origin versus all other origins. The allocation among unshocked origins remains unrestricted. At $\delta=0$, buyers share the pooled group sourcing rate. At $\delta=1$, only nonnegativity and group accounting remain.

Let $\Fcal_\delta(y)$ denote the networks satisfying the release and these sourcing restrictions. The sets are nested in $\delta$, with $\Fcal_1(y)=\Fcal(y)$ for the group release used here.

\begin{proposition}[Sharp buyer bounds]
\label{prop:binary_bounds}
Suppose $0<c_j<G_g$. Define
\begin{align*}
\rho_j&=\frac{G_g-c_j}{c_j},
&\beta_j&=\min\{1,\rho_j\},\\
a_j^-&=\min\{\bar p_g,\rho_j(1-\bar p_g)\},
&a_j^+&=\min\{1-\bar p_g,\rho_j\bar p_g\}.
\end{align*}
Under nonnegativity, group accounting, and \eqref{eq:binary_restriction}, the sharp bounds on $p_j$ are
\begin{align}
p_j^L(\delta)
&=
\bar p_g-\min\{\beta_j\delta,a_j^-\},
\label{eq:share_lower}\\
p_j^U(\delta)
&=
\bar p_g+\min\{\beta_j\delta,a_j^+\}.
\label{eq:share_upper}
\end{align}
For $c_j=G_g$, group accounting fixes $p_j=\bar p_g$. Summing the sharp supplier-block bounds gives sharp exposure bounds when the blocks are otherwise separable.
\end{proposition}

The factor $\beta_j$ limits a buyer's deviation to what the rest of its group can offset. The caps $a_j^-$ and $a_j^+$ also account for nonnegativity and the pooled sourcing share.

For target $t$, write the population identified set as $C_t(\delta)=[L_t(\delta),U_t(\delta)]$ and define its width
\begin{equation}
W_t(\delta)=U_t(\delta)-L_t(\delta).
\label{eq:assumption_precision_frontier}
\end{equation}
The width $W_t(\delta)$ is weakly increasing because relaxing the sourcing restriction expands the feasible set. Common sourcing gives $W_t(0)=0$, while $W_t(1)$ measures uncertainty under group accounting alone. The difference $W_t(1)-W_t(\delta)$ therefore measures the reduction in width supplied by the restriction.

For each supplier, the width of the buyer's sourcing share is
\begin{equation}
w_j(\delta)=\min\{\beta_j\delta,a_j^-\}
             +\min\{\beta_j\delta,a_j^+\}.
\label{eq:frontier_capacity}
\end{equation}
For $0<\bar p_g<1$, its initial slope is $2\beta_j$, with saturation points $a_j^-/\beta_j$ and $a_j^+/\beta_j$. Exposure width sums these terms with supplier input weights $c_{idj}/D_{dj}$. Appendix \ref{app:inference} compares the resulting population widths with intervals that also include sampling uncertainty.

\subsection{Breakdown points for threshold conclusions}

For a positive conclusion $c_t'f\geq\kappa_t$, define
\begin{equation}
\delta_t^+
=
\sup\{\delta\in[0,1]:L_t(\delta)\geq\kappa_t\}.
\label{eq:positive_breakdown}
\end{equation}
For a negative conclusion $c_t'f\leq\kappa_t$, define $\delta_t^-=\sup\{\delta\in[0,1]:U_t(\delta)\leq\kappa_t\}$. These quantities are the breakdown points $\delta^*$ of Section \ref{sec:completion} for positive and negative conclusions, in the sense of \citet{MastenPoirier2020}. A value of one means the cross-tab alone identifies the conclusion. A value near zero means the conclusion requires nearly common sourcing.

\begin{figure}[!htbp]
\centering
\includegraphics[width=0.98\textwidth]{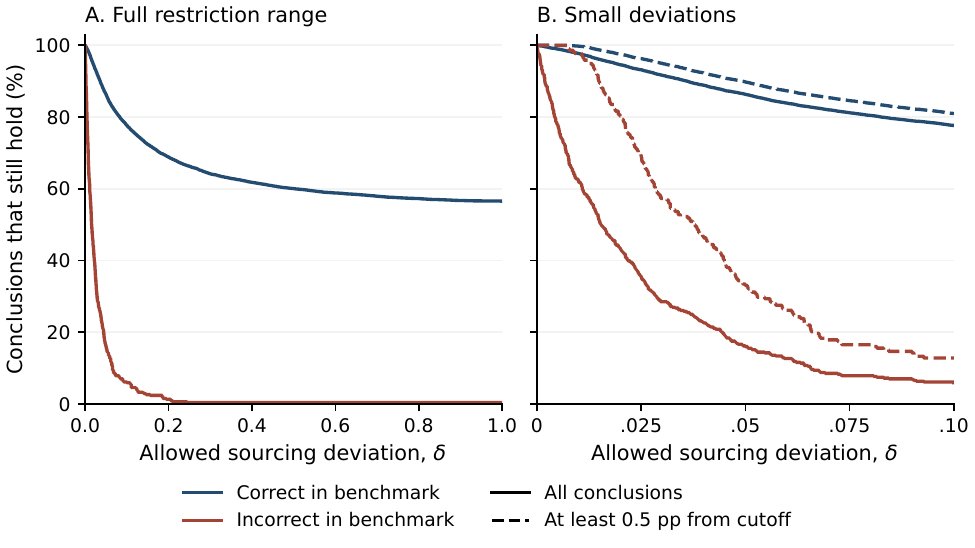}
\caption{Distribution of threshold breakdown points}
\label{fig:breakdown_distribution}
\begin{minipage}{0.95\textwidth}
\vspace{0.3em}
\footnotesize
\textit{Notes}: The curves report the share of buyer-group completion conclusions that hold for every compatible allocation at each $\delta$. Shares are calculated separately within each correctness and distance group. Conclusions cover the 2 and 3 percent thresholds in all 27 localized shock designs. Correctness is evaluated in the held-out 2005 table. Panel A shows all 12,528 conclusions. Panel B enlarges $0\leq\delta\leq0.10$: solid lines use all conclusions, and dashed lines require completed exposure at least 0.5 percentage points from the threshold. A sourcing deviation of 0.05 is five percentage points.
\end{minipage}
\end{figure}

Figure \ref{fig:breakdown_distribution} compares breakdown points for correct and incorrect classifications at the 2 and 3 percent thresholds. To assess whether the difference reflects proximity to the cutoff, Panel B also restricts the comparison to completed exposures at least 0.5 percentage points from the threshold.

The 456 incorrect conclusions have a median breakdown point of 0.015, and 15.8 percent survive a deviation of 0.05. Among the 12,072 correct conclusions, the median is one and 86.3 percent survive that deviation. At a deviation of 0.10, the corresponding shares are 5.9 and 77.7 percent. Conditioning on distance from the cutoff narrows the difference but does not remove it.

\subsection{Monitoring-set membership}

Separate exposure intervals give a sufficient condition for top-twenty membership. A buyer must be in the top twenty if at most nineteen competitors have upper bounds at or above its lower bound. For each buyer selected by the Tohoku buyer-group completion, we solve a mixed-integer program for the smallest $\delta$ at which at least twenty competitors can have higher exposure in one feasible table. This is the breakdown point for its top-twenty membership.

\begin{samepage}
Appendix Table \ref{tab:membership_breakdown} shows that membership at the bottom of the selected set is fragile. The five false inclusions have a median breakdown point of 0.0015, compared with 0.0109 among the fifteen correct selections.\par
\end{samepage}

\subsection{Stability across vintages}

How should $\delta$ be chosen? We harmonize the official 1995 and 2005 METI tables and a METI-hosted experimental 2000 table to a common 26-sector classification \citep{METI1995,METI2000Hosted,METI2005}. For each eligible supplier-destination-buyer cell, we calculate the absolute deviation from the pooled Tohoku group share. We evaluate values of $\delta$ calibrated in 1995 against 2000 and 2005, and values calibrated in 2000 against 2005. The official 1995-to-2005 comparison is primary. The experimental 2000 table was prepared for topic analysis rather than official publication.

\begin{table}[H]
\centering
\caption{Performance of sourcing restrictions across METI vintages}
\label{tab:cross_vintage}
\small\setstretch{1.05}
\begin{tabular}{@{}lrrrrrr@{}}
\toprule
\thead{Calibration--\\validation} & \thead{Quantile\\(\%)} & \thead{Deviation\\$\delta$} & \thead{Cells\\included (\%)} & \thead{Purchases\\included (\%)} & \thead{Targets\\included (\%)} & \thead{Incorrect\\classifications} \\
\midrule
1995 -- 2005 & 50 & 0.0018 & 45.1 & 66.0 & 23.6 & 8 \\
1995 -- 2005 & 75 & 0.0121 & 71.6 & 88.4 & 79.8 & 2 \\
1995 -- 2005 & 90 & 0.0404 & 88.9 & 97.5 & 99.0 & 1 \\
1995 -- 2005 & 95 & 0.0841 & 95.0 & 98.9 & 100.0 & 0 \\
\addlinespace[0.45em]
1995 -- 2000 & 90 & 0.0404 & 88.7 & 97.1 & 99.0 & 0 \\
1995 -- 2000 & 95 & 0.0841 & 95.1 & 98.9 & 100.0 & 0 \\
\addlinespace[0.45em]
2000 -- 2005 & 90 & 0.0459 & 90.2 & 97.7 & 99.0 & 0 \\
2000 -- 2005 & 95 & 0.0820 & 94.8 & 98.9 & 100.0 & 0 \\
\bottomrule
\end{tabular}
\begin{minipage}{0.97\textwidth}
\vspace{0.4em}
\footnotesize
\textit{Notes}: The allowed deviation $\delta$ is calibrated from positive-purchase buyer--supplier-industry cells across all regions and common sectors. Cell inclusion is the fraction whose Tohoku sourcing-share deviation does not exceed $\delta$, while purchase inclusion weights these cells by their supplier-specific purchases. Target inclusion instead concerns the 208 outside-Tohoku manufacturing exposures. Incorrect classifications count conclusions at the 2 and 3 percent thresholds that contradict the benchmark. The value of $\delta$ is not recalibrated in validation. The 2000 table is experimental. All comparisons are deterministic evaluations of constructed tables.
\end{minipage}
\end{table}

\begin{samepage}
Setting $\delta$ to the 1995 median deviation excludes some official 2005 targets (Table \ref{tab:cross_vintage}). When calibration uses the 90th percentile, the exposure intervals contain 99.0 percent of benchmark targets in each forward check. The 1995-to-2005 comparison has one incorrect classification. At the 95th percentile, all three comparisons include every target and make no incorrect classification at either threshold.
\end{samepage}

An exposure interval can contain the benchmark value even when some underlying transaction cells violate the sourcing restriction. At the 1995 calibration ($\delta=0.0841$, rounded), 55 of 2,621 active manufacturing-input cells for outside-Tohoku buyers violate it, and the largest deviation is 0.332. Nevertheless, one allocation satisfies the releases and this sourcing restriction while reproducing all 208 benchmark exposures. This agreement concerns direct exposure. It does not establish agreement for indirect responses. These comparisons assess stability across constructed accounts and their construction procedures. They do not independently validate sourcing restrictions in underlying transactions, or establish an appropriate $\delta$ for other countries, shocks, or classifications.

\begin{samepage}
The sourcing restriction also changes maximum regret. Holding the shocked-origin group release and monitoring set fixed, group accounting alone gives maximum regret of 18.902 points. The calibrated $\delta$ reduces it to 4.132 points. Common sourcing fixes the exposure vector and reduces maximum regret to zero, but excludes the benchmark target vector. The selected set still loses 0.217 points in the benchmark at every value of $\delta$.

\end{samepage}

\section{Inference with Estimated Releases}
\label{sec:use}

The preceding bounds treat released totals as exact. When those totals are estimated, we must also allow for sampling error. Projection inference accounts for both sources of uncertainty, whose relative importance we examine in a sampling experiment calibrated to the Japanese table.

\subsection{Sampling uncertainty under completion}

Let $f_0$ denote the population network and $y_0=Af_0$ its release. For target $t$, distinguish population exposure from the exposure implied by the completion rule when released totals are known exactly:
\begin{equation*}
\theta_t=c_t'f_0,\qquad \theta_t^m=c_t'T_m(y_0).
\end{equation*}
A completion standard error measures how sampling variation in $\widehat y$ changes $c_t'T_m(\widehat y)$ while the allocation rule is held fixed. The resulting confidence interval concerns the exposure $\theta_t^m$ implied by the completion rule. Its coverage guarantee extends to population exposure $\theta_t$ only when the completion is correct for that target or the release already identifies it.

\begin{proposition}[No uniform precision without identification]
\label{prop:no_uniform_precision}
Suppose two networks $f_0$ and $f_1$ generate the same distribution of observed releases at every sample size, but $c_t'f_0\neq c_t'f_1$. Let $C_{n,t}$ be any release-measurable random interval. If $\operatorname{diam}(C_{n,t})\xrightarrow{p}0$ under the common release law, then
\begin{equation}
\limsup_{n\to\infty}
\min_{k\in\{0,1\}}
\Pr_{f_k}\{c_t'f_k\in C_{n,t}\}
\leq \frac{1}{2}.
\label{eq:no_uniform_precision}
\end{equation}
Thus a shrinking interval cannot have uniform coverage above one half over observationally equivalent networks. Moreover, if an interval conditional on the completion rule contracts in probability to $\theta_t^m$ and $\theta_t^m\neq\theta_t$, its coverage of $\theta_t$ converges to zero.
\end{proposition}

\subsection{Simultaneous inference over targets and restrictions}

Let $y_0=Af_0$ be estimated by $\widehat y$, with simultaneous confidence region $\Ycal_n(1-\alpha)$. We retain every network compatible with this region. For each $\delta\in[0,1]$, the projected network set is
\begin{equation}
\widehat\Fcal_{n,\delta}
=
\bigcup_{y\in\Ycal_n(1-\alpha)}\Fcal_\delta(y).
\label{eq:projected_set}
\end{equation}
Let $\widehat C_{n,t}(\delta)=[\widehat L_{n,t}(\delta),\widehat U_{n,t}(\delta)]$ denote the range of target $t$ over \eqref{eq:projected_set}, and let $\widehat W_{n,t}(\delta)$ be its width.

\begin{proposition}[Simultaneous inference for the restriction path]
\label{prop:projection_coverage}
Let $\mathcal T$ be any prespecified collection of linear targets. If $\Pr\{y_0\in\Ycal_n(1-\alpha)\}\geq1-\alpha-r_n$, then
\begin{equation}
\Pr\left\{
C_t(\delta)\subseteq\widehat C_{n,t}(\delta)
\text{ for every }(t,\delta)\in\mathcal T\times[0,1]
\right\}
\geq1-\alpha-r_n.
\label{eq:network_coverage}
\end{equation}
Consequently, $W_t(\delta)\leq\widehat W_{n,t}(\delta)$ on the same event, so $\widehat W_{n,t}(\delta)$ is a simultaneous upper confidence band for the entire width curve. No further multiplicity adjustment over targets or values of $\delta$ is required once the release region is simultaneous.
\end{proposition}

The proposition specializes the projection methods in \citet{ChernozhukovHongTamer2007} and \citet{KaidoMolinariStoye2019} to a nested path of linear network restrictions. It complements confidence intervals for partially identified parameters in \citet{ImbensManski2004} and \citet{Stoye2009}.

The projected lower endpoint gives a lower confidence bound on the breakdown point for a positive conclusion:
\begin{equation}
\underline\delta_{n,t}
=
\sup\{\delta\in[0,1]:\widehat L_{n,t}(\delta)\geq\kappa_t\}.
\label{eq:breakdown_lcb}
\end{equation}

\begin{samepage}
\begin{corollary}[Simultaneous lower confidence bounds for breakdown points]
\label{cor:breakdown_inference}
Under the conditions of Proposition \ref{prop:projection_coverage},
\begin{equation}
\Pr\left\{
\underline\delta_{n,t}\leq\delta_t^+
\text{ for every }t\in\mathcal T
\right\}
\geq1-\alpha-r_n.
\label{eq:breakdown_coverage}
\end{equation}
The analogous statement holds jointly for negative conclusions using projected upper endpoints.
\end{corollary}
\end{samepage}

\subsection{A calibrated sampling experiment}

We use the common 26-sector 2005 table and its 208 outside-Tohoku manufacturing-exposure targets. We hold buyer input totals and supplier-destination-group totals fixed and estimate each stratum's Tohoku share with independent binomial samples. The experiment has 1,000 replications at counts of 500, 2,000, 10,000, and 50,000 per active stratum, independent of METI survey sample sizes.

For each buyer, we construct a confidence region covering the released shares of all inputs it purchases. The adjustment ensures joint coverage of that buyer's input shares, but does not guarantee simultaneous coverage across buyers. Proposition \ref{prop:projection_coverage} shows how a release region covering all targets would provide that stronger guarantee. Appendix \ref{app:binomial_sampling} gives the sampling details.

We compare three procedures. First, the confidence interval under common sourcing sets $\delta=0$: it allows released shares to vary within their confidence region while imposing that restriction. Second, calibration using 1995 data sets $\delta$ to the 95th percentile of observed sourcing deviations, $\delta=0.0840847$. Finally, the interval under accounting restrictions alone sets $\delta=1$. The latter two procedures project the same target-specific release region through the corresponding network set.

\begin{figure}[H]
\centering
\includegraphics[width=0.98\textwidth]{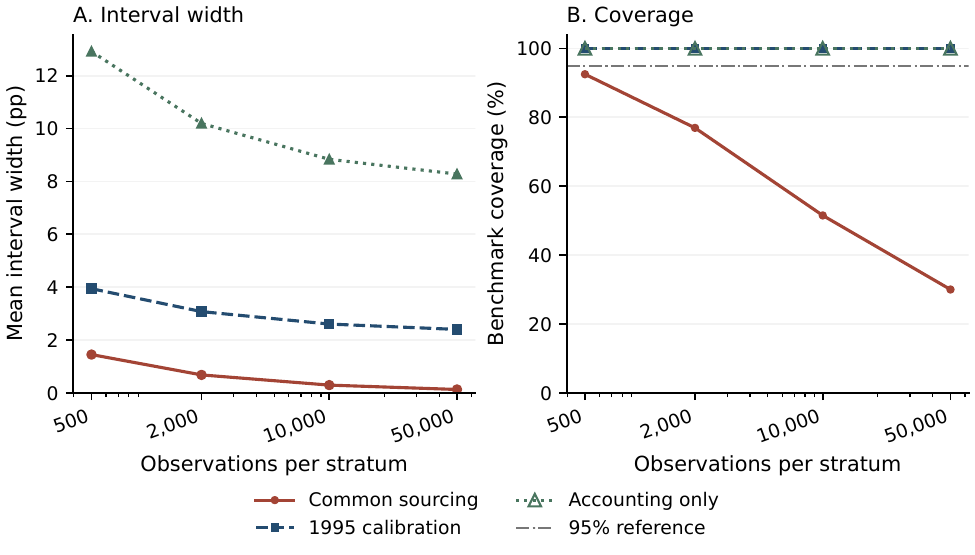}
\caption{Sampling precision and network identification}
\label{fig:assumption_precision}
\begin{minipage}{0.95\textwidth}
\vspace{0.3em}
\footnotesize\setstretch{1.1}
\textit{Notes}: Each point averages over 208 outside-Tohoku targets and 1,000 replications. Sample size is the binomial count per supplier--destination--buyer-group stratum, with purchase totals held exact. Clopper--Pearson intervals use a target-specific Bonferroni adjustment. Common sourcing, the calibrated deviation, and accounting alone set $\delta$ to 0, 0.0840847, and 1, respectively. Coverage is the percentage of target--replication pairs containing benchmark exposure, not the frequency of covering all targets jointly. Coverage under the calibrated deviation and accounting alone coincides at 100 percent.
\end{minipage}
\end{figure}

Figure \ref{fig:assumption_precision} compares the resulting interval widths and benchmark coverage as the sample grows.

\begin{samepage}
At $n=500$, the mean common-sourcing interval is 1.45 percentage points wide and covers the benchmark in 92.5 percent of target-replication pairs. At $n=50{,}000$, its width falls to 0.13 points, but benchmark coverage falls to 30.0 percent. Purchase-weighted coverage is 36.1 percent. The interval still covers the exposure implied by common sourcing in more than 99.9 percent of target-replication pairs.
\end{samepage}

At the largest sample, the interval using the calibrated $\delta$ is 2.40 percentage points wide and the interval under accounting restrictions alone is 8.28 points wide. Both contain benchmark exposure in all 208,000 target--replication pairs and imply no incorrect threshold classifications. Intervals under common sourcing imply incorrect classifications in 2.16 percent of pairs at the 2 percent threshold and 1.44 percent at the 3 percent threshold. Appendix Tables \ref{tab:assumption_precision} and \ref{tab:threshold_certification} report all sample sizes.

Appendix Table \ref{tab:inference} reports two further experiments: one estimates origin and buyer margins jointly from multinomial flows, and the other evaluates a projected lower confidence bound for the breakdown point. These coverage probabilities describe the controlled sampling designs. METI does not report the covariance needed to calculate corresponding empirical standard errors.

\section{Direct and Indirect Exposure}
\label{sec:propagation}

Do exact measurements of direct responses identify indirect exposure? Consider a fixed-coefficient linear cost model,
\begin{equation}
u=q+\Omega u,\qquad \Omega_{bs}=\frac{F^D_{sb}}{X_b},
\label{eq:linear_cost_model}
\end{equation}
where $u$ is the total unit-cost response and $q$ is an exogenous direct unit-cost disturbance. The indices $s$ and $b$ denote supplier and buyer region-industries. Domestic transactions are $F^D_{sb}$, and gross output $X_b>0$ is observed. Imported inputs enter costs at fixed foreign prices, so their price disturbance is zero. The first input-cost response is $v=\Omega q$, and the second is $\Omega v$. For a binary manufacturing shock and an outside-origin buyer, $v_b=(D_b/X_b)E_b$. Thus the denominator changes from intermediate purchases to gross output.

\subsection{Identification beyond direct responses}

An economy with three regions and one industry separates the two information requirements. Set $X_b=1$, $F^D(t)=\Omega(t)'$, $q=(1,0,0)'$, and
\begin{equation}
\Omega(t)=\frac12
\begin{pmatrix}
0.5&0.3&0.2\\
0.2&0.4+t&0.4-t\\
0.3&0.3-t&0.4+t
\end{pmatrix},\qquad -0.2\leq t\leq0.2.
\label{eq:propagation_example}
\end{equation}
Every row and column sum is $1/2$, and the entire direct-response vector is fixed at $(0.25,0.10,0.15)'$. Nevertheless,
\begin{equation}
\Omega(t)^2q=(0.0925,\ 0.075-0.025t,\ 0.0825+0.025t)'.
\label{eq:propagation_example_second}
\end{equation}
The first shock does not distinguish suppliers 2 and 3, but their different direct responses matter in the next round. A reallocation between them changes indirect exposure while preserving the release and every direct response. The aggregate remains identified: $\mathbf1'u=2$ for every $t$.

\begin{proposition}[Sharp next-round bounds conditional on measured direct responses]
\label{prop:conditional_second}
Suppose gross outputs and the complete vector $v=\Omega q$ are observed. On the nonempty accounting set augmented by this equality, the second-round response of buyer $b$ is the linear functional
\begin{equation}
(\Omega^2q)_b=\sum_s\frac{v_s}{X_b}F^D_{sb}.
\label{eq:conditional_second}
\end{equation}
Its sharp endpoints are linear programs. It is identified if and only if its coefficient belongs to the row space of the augmented release operator on the effective support.
\end{proposition}

The proof substitutes the measured $v$ and applies Lemma \ref{lem:rowspace}. Appendix \ref{app:invariant_span} gives sufficient aggregate information for several rounds.

\subsection{Second-round ambiguity in Japan}

We construct domestic inputs by allocating imports proportionally across own-region uses in the detailed 53-sector table, then aggregate to 26 sectors. Imports account for \ImportedInputPercent\ percent of retained intermediate purchases. Appendix \ref{app:network_propagation} gives the allocation and reconciliation details and establishes stability of the domestic accounting set.

The group release here observes all origin--destination--buyer-group totals for all 26 suppliers. Section \ref{sec:information} uses only Tohoku manufacturing totals. Adding $v$ discloses 234 response entries. The group release already determines 27 sums of these responses, each weighted by buyer gross output. Because of these restrictions, only 207 entries add independent information.

\begin{table}[!htbp]
\centering
\caption{Direct information and remaining second-round uncertainty}
\label{tab:network_second}
\small
\begin{tabular}{@{}lrrl@{}}
\toprule
Information & \thead{Direct width\\(pp)} & \thead{Second-round\\width (pp)} & \thead{Second-round\\bound} \\
\midrule
Domestic and buyer margins & 19.790 & 10.650 & Outer \\
Full domestic buyer-group totals & 4.418 & 2.677 & Outer \\
Group totals and measured $v$ & 0.000 & 0.360 & Sharp \\
\bottomrule
\end{tabular}
\begin{minipage}{0.95\textwidth}
\vspace{0.4em}
\footnotesize
\textit{Notes}: Mean widths across 208 outside-Tohoku buyers, in percentage points of response to a unit Tohoku manufacturing cost disturbance. All direct widths are sharp. Without measured $v$, second-round exposure is quadratic in the network. The first two rows report certified outer bounds, not sharp intervals. The final row reports sharp conditional second-round bounds. Widths use gross-output normalization and are not directly comparable to purchase-share widths.
\end{minipage}
\end{table}

Table \ref{tab:network_second} reports how the bounds narrow as information increases. Without measured direct responses, second-round exposure is quadratic in the unknown flows, and we report outer bounds. Once every direct response is observed, the problem becomes linear, yet sharp second-round width still averages \DomesticSecondMean\ percentage points and is positive for all 208 outside-Tohoku buyers (Appendix Figure \ref{fig:network_propagation}). Positive mean widths also remain for every other regional manufacturing shock (Appendix Table \ref{tab:domestic_regional_second}).

Even requiring every buyer within a group to purchase the same share from the shocked origin ($\delta=0$) leaves positive second-round width for all 208 outside-Tohoku buyers under full domestic group totals. Suppliers outside the initially shocked region can experience different direct input-cost responses. Changing which of these suppliers serves a buyer therefore changes the cost increase passed on in the next round.

\section{Conclusion}
\label{sec:conclusion}

Regional production networks are often only partially observed, but an economic decision may require much less information than full network reconstruction. In the Japanese accounts, measurements selected from an earlier vintage produce a nearly optimal benchmark monitoring set while leaving substantial maximum regret across compatible networks. In the U.S. coal application, additional shipment information establishes the optimal monitoring set even though some monitored plants' exposures remain unidentified. Evaluating releases by the decisions they support makes this distinction between benchmark performance and a decision guarantee explicit.

These conclusions are conditional on the stated accounting information and restrictions. Agreement with constructed accounts does not independently validate underlying sourcing patterns. Where sourcing restrictions remain necessary, researchers should distinguish the precision supplied by those assumptions from the precision gained by reducing sampling uncertainty. Information sufficient to identify direct responses also need not identify subsequent rounds of propagation.

\clearpage
\bibliographystyle{chicago}
\bibliography{economic_exposure_without_network_completion}

@article{TodoNakajimaMatous2015,
  author  = {Todo, Yasuyuki and Nakajima, Kentaro and Matous, Petr},
  title   = {How Do Supply Chain Networks Affect the Resilience of Firms to Natural Disasters? {Evidence} from the {Great East Japan Earthquake}},
  journal = {Journal of Regional Science},
  year    = {2015},
  volume  = {55},
  number  = {2},
  pages   = {209--229},
  doi     = {10.1111/jors.12119}
}

@article{OosterhavenBouwmeester2016,
  author  = {Oosterhaven, Jan and Bouwmeester, Maaike C.},
  title   = {A New Approach to Modeling the Impact of Disruptive Events},
  journal = {Journal of Regional Science},
  year    = {2016},
  volume  = {56},
  number  = {4},
  pages   = {583--595},
  doi     = {10.1111/jors.12262}
}

@article{ChernozhukovHongTamer2007,
  author  = {Chernozhukov, Victor and Hong, Han and Tamer, Elie},
  title   = {Estimation and Confidence Regions for Parameter Sets in Econometric Models},
  journal = {Econometrica},
  year    = {2007},
  volume  = {75},
  number  = {5},
  pages   = {1243--1284},
  doi     = {10.1111/j.1468-0262.2007.00794.x}
}

@article{KaidoMolinariStoye2019,
  author  = {Kaido, Hiroaki and Molinari, Francesca and Stoye, J{\"o}rg},
  title   = {Confidence Intervals for Projections of Partially Identified Parameters},
  journal = {Econometrica},
  year    = {2019},
  volume  = {87},
  number  = {4},
  pages   = {1397--1432},
  doi     = {10.3982/ECTA14075}
}

@article{MastenPoirier2020,
  author  = {Masten, Matthew A. and Poirier, Alexandre},
  title   = {Inference on Breakdown Frontiers},
  journal = {Quantitative Economics},
  year    = {2020},
  volume  = {11},
  number  = {1},
  pages   = {41--111},
  doi     = {10.3982/QE1288}
}

@article{Searle1965,
  author  = {Searle, S. R.},
  title   = {Additional Results Concerning Estimable Functions and Generalized Inverse Matrices},
  journal = {Journal of the Royal Statistical Society, Series B (Methodological)},
  year    = {1965},
  volume  = {27},
  number  = {3},
  pages   = {486--490},
  doi     = {10.1111/j.2517-6161.1965.tb00608.x}
}

@article{Bolker1972,
  author  = {Bolker, Ethan D.},
  title   = {Transportation Polytopes},
  journal = {Journal of Combinatorial Theory, Series B},
  year    = {1972},
  volume  = {13},
  number  = {3},
  pages   = {251--262},
  doi     = {10.1016/0095-8956(72)90060-3}
}

@misc{METI1995,
  author       = {{Ministry of Economy, Trade and Industry}},
  title        = {1995 Inter-Regional Input--Output Table},
  year         = {2001},
  howpublished = {Government of Japan},
  url          = {https://www.meti.go.jp/statistics/tyo/tiikiio/result/result_1.html},
  note = {\href{https://www.meti.go.jp/statistics/tyo/tiikiio/result/result_1.html}{Source files}}
}

@misc{METI2000Hosted,
  author       = {{Ministry of Economy, Trade and Industry}},
  title        = {2000 Experimental Inter-Regional Input--Output Table: Hosted Research Materials},
  year         = {2007},
  howpublished = {Web catalogue, Government of Japan},
  url          = {https://www.meti.go.jp/statistics/tyo/tiikiio/result/result_s1.html},
  note = {Privately constructed table, not an official {METI} publication. Catalogue updated October 1, 2007. \href{https://www.meti.go.jp/statistics/tyo/tiikiio/result/result_s1.html}{Source files}}
}

@article{West1981,
  author = {West, G. R.},
  title = {An Efficient Approach to the Estimation of Regional Input--Output Multipliers},
  journal = {Environment and Planning A},
  year = {1981},
  volume = {13},
  number = {7},
  pages = {857--867},
  doi = {10.1068/a130857}
}

@article{JiangEtAl2010,
  author = {Jiang, Xuemei and Dietzenbacher, Erik and Los, Bart},
  title = {Targeting the Collection of Superior Data for the Estimation of the Intermediate Deliveries in Regional Input--Output Tables},
  journal = {Environment and Planning A},
  year = {2010},
  volume = {42},
  number = {10},
  pages = {2508--2526},
  doi = {10.1068/a4316}
}

@article{AnandCraigVonPeter2015,
  author = {Anand, Kartik and Craig, Ben and von Peter, Goetz},
  title = {Filling in the Blanks: Network Structure and Interbank Contagion},
  journal = {Quantitative Finance},
  year = {2015},
  volume = {15},
  number = {4},
  pages = {625--636},
  doi = {10.1080/14697688.2014.968195}
}

@article{MastromatteoEtAl2012,
  author = {Mastromatteo, Iacopo and Zarinelli, Elia and Marsili, Matteo},
  title = {Reconstruction of Financial Networks for Robust Estimation of Systemic Risk},
  journal = {Journal of Statistical Mechanics: Theory and Experiment},
  year = {2012},
  volume = {2012},
  number = {3},
  pages = {P03011},
  doi = {10.1088/1742-5468/2012/03/P03011},
  url = {https://arxiv.org/abs/1109.6210}
}

@techreport{METIConstruction2010,
  author = {{Ministry of Economy, Trade and Industry}},
  title = {2005 Inter-Regional Input--Output Table: Construction Report},
  institution = {Government of Japan},
  year = {2010},
  month = {March},
  note = {In Japanese. \href{https://www.e-stat.go.jp/stat-search/file-download?fileKind=2&statInfId=000020467388}{Construction report}},
  url = {https://www.e-stat.go.jp/stat-search/file-download?fileKind=2&statInfId=000020467388}
}

@article{TianKannan2006,
  author = {Tian, Jianjun Paul and Kannan, D.},
  title = {Lumpability and Commutativity of {Markov} Processes},
  journal = {Stochastic Analysis and Applications},
  year = {2006},
  volume = {24},
  number = {3},
  pages = {685--702},
  doi = {10.1080/07362990600632045}
}

@techreport{BartolucciEtAl2024,
  author = {Bartolucci, Silvia and Caccioli, Fabio and Caravelli, Francesco and Vivo, Pierpaolo},
  title = {Upstreamness and Downstreamness in Input-Output Analysis from Local and Aggregate Information},
  institution = {Systemic Risk Centre, London School of Economics and Political Science},
  type = {Discussion Paper},
  number = {128},
  year = {2024},
  url = {https://www.systemicrisk.ac.uk/sites/default/files/2024-03/DP-128.pdf}
}

@book{NRC2007Coal,
  author = {{National Research Council}},
  title = {Coal: Research and Development to Support National Energy Policy},
  year = {2007},
  publisher = {The National Academies Press},
  address = {Washington, DC},
  doi = {10.17226/11977},
  url = {https://www.nationalacademies.org/read/11977/chapter/7}
}

@article{Manski2007Regret,
  author = {Manski, Charles F.},
  title = {Minimax-Regret Treatment Choice with Missing Outcome Data},
  journal = {Journal of Econometrics},
  year = {2007},
  volume = {139},
  number = {1},
  pages = {105--115},
  doi = {10.1016/j.jeconom.2006.06.006}
}

@article{Stoye2007Regret,
  author = {Stoye, J{\"o}rg},
  title = {Minimax Regret Treatment Choice with Incomplete Data and Many Treatments},
  journal = {Econometric Theory},
  year = {2007},
  volume = {23},
  number = {1},
  pages = {190--199},
  doi = {10.1017/S0266466607070089}
}

@article{DuncanDavis1953,
  author  = {Duncan, Otis Dudley and Davis, Beverly},
  title   = {An Alternative to Ecological Correlation},
  journal = {American Sociological Review},
  year    = {1953},
  volume  = {18},
  number  = {6},
  pages   = {665--666},
  doi     = {10.2307/2088122}
}

@article{Moses1955,
  author  = {Moses, Leon N.},
  title   = {The Stability of Interregional Trading Patterns and Input-Output Analysis},
  journal = {American Economic Review},
  year    = {1955},
  volume  = {45},
  number  = {5},
  pages   = {803--826},
  url     = {https://www.jstor.org/stable/1821380}
}

@article{IalongoEtAl2022,
  author  = {Ialongo, Leonardo Niccol{\`o} and de Valk, Camille and Marchese, Emiliano and Jansen, Fabian and Zmarrou, Hicham and Squartini, Tiziano and Garlaschelli, Diego},
  title   = {Reconstructing Firm-Level Interactions in the {Dutch} Input-Output Network from Production Constraints},
  journal = {Scientific Reports},
  year    = {2022},
  volume  = {12},
  pages   = {11847},
  doi     = {10.1038/s41598-022-13996-3}
}

@techreport{CensusCFS2017Methodology,
  author      = {{U.S. Census Bureau}},
  title       = {2017 Commodity Flow Survey Methodology},
  institution = {U.S. Census Bureau},
  year        = {2020},
  type        = {Technical Documentation},
  doi         = {10.21949/1523073},
  url         = {https://www2.census.gov/programs-surveys/cfs/technical-documentation/methodology/2017cfsmethodology.pdf}
}

@techreport{CensusCFS2017Questionnaire,
  author      = {{U.S. Census Bureau}},
  title       = {2017 Commodity Flow Survey Instruction Guide},
  institution = {U.S. Census Bureau},
  year        = {2016},
  type        = {Form {CFS-1100}},
  note        = {Revised October 5, 2016, for the 2017 survey},
  url         = {https://www2.census.gov/programs-surveys/cfs/technical-documentation/questionnaires/CFS-1100_17.pdf}
}

@article{AcemogluEtAl2012,
  author  = {Acemoglu, Daron and Carvalho, Vasco M. and Ozdaglar, Asuman and Tahbaz-Salehi, Alireza},
  title   = {The Network Origins of Aggregate Fluctuations},
  journal = {Econometrica},
  year    = {2012},
  volume  = {80},
  number  = {5},
  pages   = {1977--2016},
  doi     = {10.3982/ECTA9623}
}

@article{BaqaeeFarhi2019,
  author  = {Baqaee, David Rezza and Farhi, Emmanuel},
  title   = {The Macroeconomic Impact of Microeconomic Shocks: Beyond {Hulten}'s Theorem},
  journal = {Econometrica},
  year    = {2019},
  volume  = {87},
  number  = {4},
  pages   = {1155--1203},
  doi     = {10.3982/ECTA15202}
}

@article{BoehmEtAl2019,
  author  = {Boehm, Christoph E. and Flaaen, Aaron and Pandalai-Nayar, Nitya},
  title   = {Input Linkages and the Transmission of Shocks: Firm-Level Evidence from the 2011 {T\={o}hoku} Earthquake},
  journal = {Review of Economics and Statistics},
  year    = {2019},
  volume  = {101},
  number  = {1},
  pages   = {60--75},
  doi     = {10.1162/rest_a_00750}
}

@article{BoeroEtAl2018,
  author  = {Boero, Riccardo and Edwards, Brian K. and Rivera, Michael K.},
  title   = {Regional Input--Output Tables and Trade Flows: An Integrated and Interregional Non-Survey Approach},
  journal = {Regional Studies},
  year    = {2018},
  volume  = {52},
  number  = {2},
  pages   = {225--238},
  doi     = {10.1080/00343404.2017.1286009}
}

@article{CaliendoEtAl2018,
  author  = {Caliendo, Lorenzo and Parro, Fernando and Rossi-Hansberg, Esteban and Sarte, Pierre-Daniel},
  title   = {The Impact of Regional and Sectoral Productivity Changes on the {U}.{S}. Economy},
  journal = {Review of Economic Studies},
  year    = {2018},
  volume  = {85},
  number  = {4},
  pages   = {2042--2096},
  doi     = {10.1093/restud/rdx082}
}

@article{CarvalhoEtAl2021,
  author  = {Carvalho, Vasco M. and Nirei, Makoto and Saito, Yukiko U. and Tahbaz-Salehi, Alireza},
  title   = {Supply Chain Disruptions: Evidence from the {Great East Japan Earthquake}},
  journal = {Quarterly Journal of Economics},
  year    = {2021},
  volume  = {136},
  number  = {2},
  pages   = {1255--1321},
  doi     = {10.1093/qje/qjaa044}
}

@book{MillerBlair2009,
  author    = {Miller, Ronald E. and Blair, Peter D.},
  title     = {Input-Output Analysis: Foundations and Extensions},
  publisher = {Cambridge University Press},
  year      = {2009},
  edition   = {2nd},
  address   = {Cambridge},
  doi       = {10.1017/CBO9780511626982}
}

@article{CrossManski2002,
  author  = {Cross, Philip J. and Manski, Charles F.},
  title   = {Regressions, Short and Long},
  journal = {Econometrica},
  year    = {2002},
  volume  = {70},
  number  = {1},
  pages   = {357--368},
  doi     = {10.1111/1468-0262.00279}
}

@article{ImbensManski2004,
  author  = {Imbens, Guido W. and Manski, Charles F.},
  title   = {Confidence Intervals for Partially Identified Parameters},
  journal = {Econometrica},
  year    = {2004},
  volume  = {72},
  number  = {6},
  pages   = {1845--1857},
  doi     = {10.1111/j.1468-0262.2004.00555.x}
}

@article{Stoye2009,
  author  = {Stoye, J{\"o}rg},
  title   = {More on Confidence Intervals for Partially Identified Parameters},
  journal = {Econometrica},
  year    = {2009},
  volume  = {77},
  number  = {4},
  pages   = {1299--1315},
  doi     = {10.3982/ECTA7347}
}

@article{CanningWang2005,
  author  = {Canning, Patrick and Wang, Zhi},
  title   = {A Flexible Mathematical Programming Model to Estimate Interregional Input--Output Accounts},
  journal = {Journal of Regional Science},
  year    = {2005},
  volume  = {45},
  number  = {3},
  pages   = {539--563},
  doi     = {10.1111/j.0022-4146.2005.00383.x}
}

@misc{METI2005,
  author       = {{Ministry of Economy, Trade and Industry}},
  title        = {2005 Inter-Regional Input--Output Table},
  year         = {2010},
  howpublished = {Government of Japan},
  url          = {https://www.meti.go.jp/english/statistics/tyo/tiikiio/index.html},
  note = {\href{https://www.meti.go.jp/english/statistics/tyo/tiikiio/index.html}{Source files}}
}

\clearpage
\appendix
\onehalfspacing
\numberwithin{equation}{section}
\numberwithin{table}{section}
\numberwithin{figure}{section}

\begin{center}
{\Large\textbf{Online Appendix}}\par
\vspace{0.4em}
{\large\textit{Decision-Relevant Information in Partially Observed Production Networks}}\par
\end{center}
\vspace{1em}

This appendix provides the proofs in Appendix \ref{app:proofs}, data construction and completion methods in Appendix \ref{app:data}, and supplementary Japanese results in Appendix \ref{app:supplementary}. Appendix \ref{app:network_measurement} derives the information counts, measurement allocation, and monitoring guarantees. Appendix \ref{app:inference} describes the sampling experiments, and Appendix \ref{app:network_propagation} develops the propagation results. Coal data construction and additional comparisons appear in the accompanying supplement, \textit{Coal Deliveries: Construction and Supplementary Results}.

\section{Proofs and Identification}
\label{app:proofs}

\subsection{Proof of Lemma \ref{lem:rowspace}}

Let $f^\circ$ be a relative-interior point of the minimal face containing $\Fcal(y)$, and restrict attention to its effective support $\mathcal E$. Every sufficiently small perturbation $v$ with $A_{\mathcal E}v=0$ remains feasible in both directions around $f^\circ$. If $c'f$ is constant, then $c_{\mathcal E}'v=0$ for every $v\in\ker(A_{\mathcal E})$. Hence $c_{\mathcal E}\in\ker(A_{\mathcal E})^\perp=\Row(A_{\mathcal E})$. Conversely, if $c_{\mathcal E}=A_{\mathcal E}'\lambda$, then $c_{\mathcal E}'f_{\mathcal E}=\lambda'A_{\mathcal E}f_{\mathcal E}=\lambda'y$ for every feasible $f$. Coordinates outside $\mathcal E$ are always zero and do not affect the functional. \hfill$\square$

\subsection{Proof of Lemma \ref{lem:additive}}

The row and column margin operator is the node-edge incidence map of the effective bipartite graph, up to an orientation. Its row space consists of edge coefficients formed by an origin potential plus a buyer potential. Hence \eqref{eq:additive_coefficients} is sufficient. The orthogonal complement of this row space is the graph's cycle space. A coefficient is in the row space if and only if it has zero inner product with every signed cycle vector, which is the zero alternating-sum condition. On each connected component, choose a reference origin, set its potential to zero, and recover all other potentials along graph paths. Cycle consistency makes the recovered values path-independent. \hfill$\square$

For $C_{rk}=q_ra_k$ on complete support, the alternating sum on a four-cycle equals \eqref{eq:rank_one_cycle}. It vanishes for every $r,r',k,k'$ only if every origin difference in $q$ is zero or every buyer difference in $a$ is zero.

\subsection{Proof of Proposition \ref{prop:target_family_rank}}

An augmented operator with additional rows $G$ identifies every $c\in\Ccal$ if and only if $\Ccal\subseteq\Row(A)+\Row(G)$. The unidentified part of $\Ccal$ has dimension $\dim(\Ccal)-\dim(\Ccal\cap\Row(A))$. Each independent added row spans at most one of these missing directions, while choosing rows that span all of them attains the bound. \hfill$\square$

For complete support, decompose each shock and incidence vector into constant and centered components. Terms containing a constant component lie in the margin row space. Bases for centered shock and incidence spaces generate linearly independent tensor products. If their dimensions are $d_Q$ and $d_A$, the interaction rank is $d_Qd_A$.

\subsection{Proof of Proposition \ref{prop:binary_bounds}}

Write $x_j=c_j(\bar p_g+u)$. Group accounting requires the other buyers, with total mass $G_g-c_j$, to have purchase-weighted mean deviation $-u/\rho_j$. For a downward move $u=-a$, feasibility requires $a\leq\delta$, $a\leq\rho_j\delta$, $a\leq\bar p_g$, and $a\leq\rho_j(1-\bar p_g)$. Collecting these restrictions gives $a\leq\min\{\beta_j\delta,a_j^-\}$. For an upward move they give $a\leq\min\{\beta_j\delta,a_j^+\}$.

Each endpoint is attainable. Assign the required common offset to the other buyers, then distribute residual slack without violating their interval or capacity bounds. The four inequalities are exactly the conditions that the required offset lies in the sum of those feasible intervals. \hfill$\square$

\subsection{Proof of Proposition \ref{prop:no_uniform_precision}}

Let $\theta_k=c_t'f_k$ and $d=|\theta_1-\theta_0|>0$. Because the observed release has the same law under $f_0$ and $f_1$, the random interval $C_{n,t}$ also has the same law. On the event $\operatorname{diam}(C_{n,t})<d$, the interval cannot contain both $\theta_0$ and $\theta_1$. Hence
\begin{equation*}
\Pr_{f_0}\{\theta_0\in C_{n,t}\}
+
\Pr_{f_1}\{\theta_1\in C_{n,t}\}
\leq
1+\Pr\{\operatorname{diam}(C_{n,t})\geq d\}.
\end{equation*}
The last probability converges to zero, so the smaller coverage probability has limit superior no greater than one half. For the second statement, choose $\epsilon<|\theta_t^m-\theta_t|/2$. If the conditional interval lies inside an $\epsilon$-neighborhood of $\theta_t^m$, it cannot contain $\theta_t$. Contraction to $\theta_t^m$ makes the probability of the complementary event converge to zero. \hfill$\square$

\subsection{Proof of Proposition \ref{prop:projection_coverage}}

On the event $y_0\in\Ycal_n$, every $f\in\Fcal_\delta(y_0)$ belongs to $\widehat\Fcal_{n,\delta}$ for every $\delta\in[0,1]$. Applying every target functional gives $C_t(\delta)\subseteq\widehat C_{n,t}(\delta)$ for all $(t,\delta)$ on the same event. Set inclusion also gives the width inequality. The event does not depend on the target coefficient or on $\delta$. \hfill$\square$

\subsection{Proof of Corollary \ref{cor:breakdown_inference}}

On the release-coverage event, the projected lower endpoint is weakly below the population lower endpoint for every target and every $\delta$. Any $\delta$ satisfying $\widehat L_{n,t}(\delta)\geq\kappa_t$ also satisfies $L_t(\delta)\geq\kappa_t$. Taking suprema gives $\underline\delta_{n,t}\leq\delta_t^+$ simultaneously for all $t\in\mathcal T$. Proposition \ref{prop:projection_coverage} supplies its probability. The negative case uses upper endpoints. \hfill$\square$

\subsection{Sourcing reallocations and exposure ambiguity}
\label{app:network_geometry}

For a fixed row-and-column-margin set, let $W(C)$ denote the sharp width of a linear functional with coefficient matrix $C$. Write $q(\lambda)=\bar q\mathbf1+\lambda\widetilde q$ and $a(\eta)=\bar a\mathbf1+\eta\widetilde a$, where $\lambda,\eta\geq0$. Then
\begin{equation}
W(q(\lambda)a(\eta)')=\lambda\eta W(\widetilde q\widetilde a').
\end{equation}
Expanding the product gives constant, supplier-only, and buyer-only terms identified by margins. Only the final interaction varies. This implication of linear estimability and transportation geometry \citep{Searle1965,Bolker1972} holds with margins, support, and target normalization fixed. Geographic concentration alone cannot order widths when these quantities differ.

The same support can give different exposure widths. Consider a complete two-by-two table with unit row and column margins, and fix $a=(1,0)'$. With $q=(1,1)'$, the target is constant. Changing the shock to $q=(1,0)'$ gives width one. Adding a disconnected, unshocked two-by-two block increases cycle dimension without changing the latter width. Expanding support while holding the target, margins, and all other restrictions fixed weakly expands the target set.

\section{Data and Completion Methods}
\label{app:data}

\subsection{Source tables and common classification}

The official 2005 29-sector workbook supplies buyer sector, supplier sector, source region, using region, and transaction value. We parse these fields into $F_{irdj}$ and treat structural blanks as zeros. Section \ref{sec:japan} describes the construction, import convention, and nonnegative cleaning. The table contains nine regions, 16 manufacturing suppliers, and 232 outside-origin targets. The common 26-sector panel retains the nine regions and has 14 manufacturing suppliers and 208 outside-origin targets.

We construct the common panel from three detailed source tables. The official 1995 workbook contains a dense 46-sector matrix. The METI-hosted experimental 2000 archive contains a 52-sector table in \texttt{S052\_XIJ.TXT}. We read it with the CP932 encoding, and each row supplies nine destination values. The detailed 2005 workbook contains a 53-sector table in long form. The accompanying crosswalk records every source code, source name, common code, and inclusion indicator for all three vintages. Reuse and recycling is excluded from both the supplier and buyer dimensions in 2000 and 2005 because it has no 1995 counterpart. All 46 source sectors are retained in 1995, 51 of 52 in 2000, and 52 of 53 in 2005.

\begin{table}[!htbp]
\centering\small
\caption{Source-to-common-sector reconciliation}
\label{tab:vintage_reconciliation}
\begin{tabular}{@{}rrrrrrr@{}}
\toprule
Year & \thead{Source\\sectors} & \thead{Retained\\sectors} & \thead{Source input\\total} & \thead{Common-panel\\input total} & \thead{Excluded\\(\%)} & \thead{Maximum\\OD error} \\
\midrule
1995 & 46 & 46 & 422,555,273 & 422,555,273 & 0.000 & 0 \\
2000 & 52 & 51 & 429,772,462 & 427,153,225 & 0.609 & 0 \\
2005 & 53 & 52 & 456,185,644 & 455,065,487 & 0.246 & 0 \\
\bottomrule
\end{tabular}
\begin{minipage}{0.97\textwidth}
\vspace{0.4em}\footnotesize
\textit{Notes}: Totals and maximum absolute origin--destination errors use each file's stored monetary units and signed transactions before nonnegative cleaning. Each reconciliation compares a vintage with its own source. The cross-vintage exercises use normalized shares. Excluded shares count all cells removed when reuse and recycling is excluded from either dimension. Every retained source transaction is assigned exactly once. The 2000 vintage is experimental.
\end{minipage}
\end{table}

Table \ref{tab:vintage_reconciliation} verifies that aggregation preserves the included source total and all 81 origin--destination totals in every vintage. The nonnegative 2005 table used for direct exposure contains 455,065,764 million yen, compared with 455,065,487 before cleaning. Negative source entries can offset other entries during aggregation, so they differ from the negative aggregated cells removed in cleaning.

\subsection{Buyer groups}

Table \ref{tab:buyer_groups} gives the prespecified Goods, Market, and Local groups in both classifications. The common classification combines services differently from the original 29-sector table. In particular, common-sector 25, Other services, belongs to Market, while common-sector 26, Others, belongs to Local. In the 29-sector table, Personal services and Others both belong to Local.

\begin{table}[!htbp]
\centering\small
\caption{Buyer-group assignments in the two sector classifications}
\label{tab:buyer_groups}
\begin{tabular}{@{}l>{\raggedright\arraybackslash}p{6.3cm}>{\raggedright\arraybackslash}p{6.0cm}@{}}
\toprule
Buyer group & Original 29-sector codes & Common 26-sector codes \\
\midrule
Goods & 10, 20, 30, 40, 50, 60, 70, 80, 90, 100, 110, 120, 130, 140, 150, 160, 170, 180 & 1, 2, 3, 4, 5, 6, 7, 8, 9, 10, 11, 12, 13, 14, 15, 16 \\
Market & 200, 210, 220, 230, 240, 270 & 18, 19, 20, 21, 22, 25 \\
Local & 190, 250, 260, 280, 290 & 17, 23, 24, 26 \\
\bottomrule
\end{tabular}
\begin{minipage}{0.95\textwidth}
\vspace{0.4em}\footnotesize
\textit{Notes}: Codes cover all 29 original sectors and all 26 common sectors, each assigned once. The replication files give the corresponding full sector names and source-to-common crosswalks.
\end{minipage}
\end{table}

\subsection{Comparability of transaction records}

Every additional release must aggregate the same producer-to-user transaction as the target flow. Freight records often end at a warehouse, distribution center, or port rather than the ultimate input user \citep{CensusCFS2017Methodology,CensusCFS2017Questionnaire}. If inbound and outbound records cannot be linked, the same shipment totals can be consistent with different matches between producers and ultimate users. The two-by-two swap in \eqref{eq:rank_one_cycle} changes those matches and hence exposure while preserving every leg total. Precise leg data therefore do not identify ultimate exposure without chain identifiers, ultimate-use destinations, or justified routing restrictions.

Payment and wholesale records likewise require comparable statistical units, valuations, populations, and timing to measure the producer-to-user transaction entering exposure.

\subsection{Completion algorithms and computation}
\label{app:computation}

Gravity-RAS starts from $P_{rd}(h,\gamma)=\exp\{h\mathbf1(r=d)-\gamma d_{rd}\}$. Here $d_{rd}$ is great-circle distance in miles, calculated from the regional reference coordinates supplied in the code. Iterative proportional fitting balances this seed to the supplier's origin and destination totals. The resulting origin--destination amounts are allocated across destination buyers in proportion to their purchases of that supplier's input.

Let $K_{ird}=\sum_jF_{irdj}$, $M_i=\sum_{r,d}K_{ird}$, and let $\widehat K_i(h,\gamma)$ denote the balanced candidate. We choose $h\in\{0,1,2,3\}$ and $\gamma\in\{0,0.0005,\ldots,0.006\}$ to minimize
\begin{equation}
\frac{1}{\sum_{i\in\mathcal N}M_i}
\sum_{i\in\mathcal N}M_i\frac{1}{81}
\sum_{r,d}\left\{\frac{\widehat K_{ird}(h,\gamma)}{M_i}
-\frac{K_{ird}}{M_i}\right\}^2,
\end{equation}
where $\mathcal N$ contains the thirteen nonmanufacturing suppliers. This calibration uses their bilateral flows in addition to baseline margins and selects $h=3$ and $\gamma=0.002$ per mile. Manufacturing bilateral flows are withheld for all twenty-seven shock designs.

As an accounting check, all four methods reproduce total supplier-family input shares, including imported inputs, to numerical precision.

For the 12,528 buyer-group conclusions at the 2 and 3 percent thresholds, bisection finds the largest $\delta$ under which the sharp endpoint remains on the stated side of the threshold. Appendix \ref{app:supplementary} describes the monitoring membership program.

\section{Supplementary Japanese Results}
\label{app:supplementary}

\subsection{Conditions associated with completion error}

\begin{figure}[!htbp]
\centering
\includegraphics[width=0.98\textwidth]{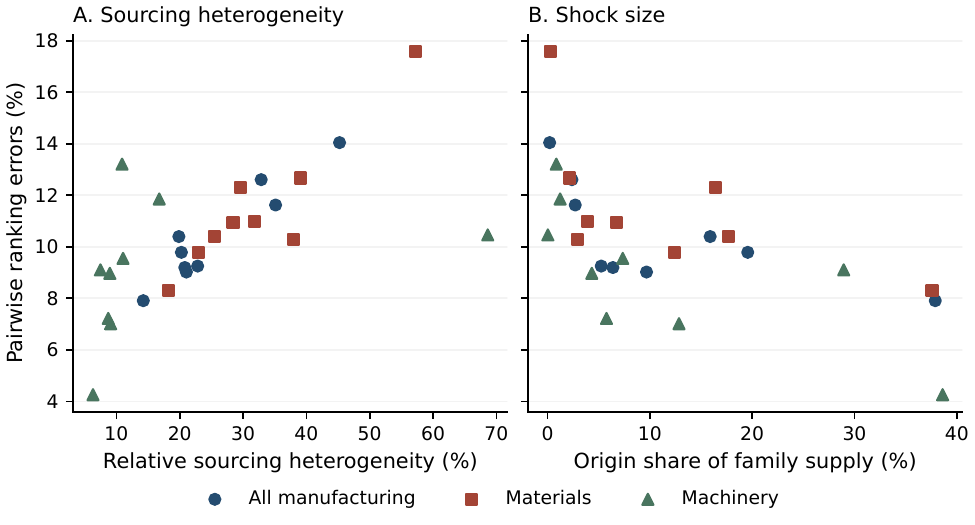}
\caption{Economic conditions associated with completion error}
\label{fig:completion_conditions}
\begin{minipage}{0.94\textwidth}
\vspace{0.3em}
\footnotesize
\textit{Notes}: Each point is one origin-region and supplier-family design for the buyer-group completion. Sourcing heterogeneity is the purchase-weighted mean absolute deviation from the pooled group sourcing share, divided by the average pooled share. Shock size is the shocked region's share of national supply in the supplier family. The figure contains 27 designs.
\end{minipage}
\end{figure}

Pairwise ranking errors rise with relative sourcing heterogeneity and are more frequent for small regional suppliers (Figure \ref{fig:completion_conditions}). The latter association is consistent with a weaker exposure signal as well as sourcing heterogeneity. Both measures were defined before examining errors. The associations are descriptive and cover 27 correlated shock designs.

\subsection{Monitoring capacity and supplier families}

Buyer-group and collapsed-bilateral completion rank first and second in mean overlap at monitoring capacities of ten, twenty, and thirty (Table \ref{tab:capacity_sensitivity}). Gravity and proportional completion exchange positions as capacity changes, but both recover little more than half of the benchmark set even at thirty.

\begin{table}[!htbp]
\centering
\caption{Monitoring performance at alternative capacities}
\label{tab:capacity_sensitivity}
\small
\begin{tabular}{@{}lrrr@{}}
\toprule
Method & $K=10$ & $K=20$ & $K=30$ \\
\midrule
Proportional margins & 36.7 & 50.0 & 57.7 \\
Gravity-RAS margins & 41.5 & 49.8 & 57.4 \\
Collapsed bilateral & 66.3 & 70.2 & 75.9 \\
Buyer-group bilateral & 73.0 & 79.8 & 82.0 \\
\bottomrule
\end{tabular}
\begin{minipage}{0.9\textwidth}
\vspace{0.4em}
\footnotesize
\textit{Notes}: Entries are mean overlap percentages across the 27 localized shock designs. $K$ is the number of monitored buyers, fixed before comparing completions.
\end{minipage}
\end{table}

Machinery has the highest relative exposure score in Table~\ref{tab:family_results}, at 97.1 percent of the optimum.

\begin{table}[!htbp]
\centering
\caption{Buyer-group completion performance by supplier family}
\label{tab:family_results}
\small
\begin{tabular}{@{}lrrr@{}}
\toprule
Supplier family & \thead{Pairwise errors\\(\%)} & \thead{Top-20 overlap\\(\%)} & \thead{Relative score\\(\%)} \\
\midrule
All manufacturing & 10.4 & 76.7 & 91.1 \\
Materials & 11.5 & 73.3 & 89.4 \\
Machinery & 9.1 & 89.4 & 97.1 \\
\bottomrule
\end{tabular}
\begin{minipage}{0.9\textwidth}
\vspace{0.4em}
\footnotesize
\textit{Notes}: Entries average over the nine origin-region shocks within each supplier family. Relative score is total benchmark exposure in the selected twenty divided by the optimum benchmark total, including contributions from selected buyers outside the benchmark top twenty.
\end{minipage}
\end{table}

\subsection{Sensitivity of monitoring-set membership}

Binary variables identify the competitors that overtake each selected buyer in the membership calculation of Section \ref{sec:sensitivity}. Group accounting holds jointly within every supplier--destination--group block, giving sharp breakdown points up to the numerical ranking tolerance.

Table \ref{tab:membership_breakdown} orders buyers by the completed ranking. Even correct selections near the cutoff can be fragile: Kanto other manufacturing is in the benchmark top twenty but can fall outside it at $\delta=0.0002$.

\begin{table}[H]
\centering
\caption{Breakdown points for membership in the Tohoku monitoring set}
\label{tab:membership_breakdown}
\small
\begin{tabular}{@{}r>{\raggedright\arraybackslash}p{5.5cm}rrrc@{}}
\toprule
Rank & Buyer & \thead{Completed\\(\%)} & \thead{Benchmark\\(\%)} & \thead{Breakdown\\$\delta^*$} & \thead{Benchmark\\top 20} \\
\midrule
1 & Hokkaido, Electrical machinery & 4.71 & 4.79 & 0.0324 & Yes \\
2 & Kanto, Electrical machinery & 4.47 & 4.33 & 0.0260 & Yes \\
3 & Kanto, Precision instruments & 3.91 & 3.86 & 0.0159 & Yes \\
4 & Kanto, Timber and furniture & 3.81 & 4.16 & 0.0156 & Yes \\
5 & Hokkaido, Precision instruments & 3.80 & 2.97 & 0.0157 & Yes \\
6 & Kanto, Non-ferrous metal products & 3.53 & 5.43 & 0.0123 & Yes \\
7 & Kanto, Transportation equipment & 3.37 & 3.57 & 0.0164 & Yes \\
8 & Shikoku, Precision instruments & 3.25 & 3.09 & 0.0109 & Yes \\
9 & Kanto, General machinery & 3.09 & 3.27 & 0.0104 & Yes \\
10 & Kanto, Pulp and paper products & 3.01 & 3.01 & 0.0078 & Yes \\
11 & Hokkaido, Plastic products & 2.87 & 2.21 & 0.0059 & No \\
12 & Shikoku, Electrical machinery & 2.69 & 2.67 & 0.0073 & Yes \\
13 & Kinki, Pulp and paper products & 2.68 & 3.58 & 0.0050 & Yes \\
14 & Kanto, Construction & 2.50 & 2.43 & 0.0055 & Yes \\
15 & Hokkaido, Pulp and paper products & 2.48 & 1.74 & 0.0038 & No \\
16 & Kanto, Textile products & 2.33 & 1.74 & 0.0015 & No \\
17 & Kyushu, Electrical machinery & 2.27 & 2.27 & 0.0013 & No \\
18 & Hokkaido, Chemical products & 2.17 & 2.53 & 0.0004 & Yes \\
19 & Chugoku, Electrical machinery & 2.17 & 1.61 & 0.0004 & No \\
20 & Kanto, Other manufacturing & 2.15 & 2.69 & 0.0002 & Yes \\
\bottomrule
\end{tabular}
\begin{minipage}{0.97\textwidth}
\vspace{0.4em}
\footnotesize
\textit{Notes}: The point ranking uses the buyer-group completion for the Tohoku manufacturing shock. The breakdown point $\delta^*$ is the smallest sourcing-share deviation at which at least twenty competitors can jointly exceed the selected buyer. The final column evaluates membership against the benchmark. Sector names are shortened, with full names in the replication crosswalk.
\end{minipage}
\end{table}

\section{Release Design and Monitoring Guarantees}
\label{app:network_measurement}

\subsection{Independent information in successive releases}

With nine origins and $9J$ buyers, where $J$ is the number of buyer industries, the margins leave $8(9J-1)$ independent directions in which a supplier block's flows can vary. This count is calculated before allowing for flows that the accounts force to zero. A shocked-origin release to $H$ buyer categories in each destination has $9H$ raw cells, but its sum is already observed in $o_{iT}$. It therefore adds $9H-1$ independent rows on complete support. Summing over suppliers gives Table~\ref{tab:network_release_counts}. Costs hold this classification-based template fixed across vintages.

In 2005, accounting-forced zeros reduce the half-budget menu's effective rank from 1,606 to 1,579 and the full menu's rank from 3,262 to 2,939. The half-budget menu thus uses 53.7 percent of the full menu's effective rank. 

\begin{table}[!htbp]
\centering\small
\caption{Information counts relative to national supplier and buyer margins}
\label{tab:network_release_counts}
\begin{tabular}{@{}rlrrr@{}}
\toprule
Sectors & Release & \thead{Cumulative\\rank} & \thead{Incremental\\rank} & \thead{Raw\\cells} \\
\midrule
26 & Margins & 0 & 0 & 0 \\
26 & Destination & 112 & 112 & 126 \\
26 & Buyer group & 364 & 252 & 378 \\
26 & Buyer industry & 3,262 & 2,898 & 3,276 \\
\addlinespace[0.45em]
29 & Margins & 0 & 0 & 0 \\
29 & Destination & 128 & 128 & 144 \\
29 & Buyer group & 416 & 288 & 432 \\
29 & Buyer industry & 4,160 & 3,744 & 4,176 \\
\bottomrule
\end{tabular}
\begin{minipage}{0.94\textwidth}
\vspace{0.3em}\footnotesize
\textit{Notes}: Counts cover 14 manufacturing suppliers in the 26-sector panel and 16 in the 29-sector table. Ranks use the complete classification template. Incremental rank is relative to the preceding nested level, and cumulative rank is relative to baseline margins. Raw cells count the displayed shocked-origin cross-tab, not redundant rest-of-origin complements. Neither rank nor raw cells is a measured confidentiality cost.
\end{minipage}
\end{table}

Full shocked-origin buyer disclosure identifies this shock's exposures but leaves allocations among unshocked origins unresolved.

\subsection{Exact allocation within the menu}
\label{app:measurement_allocation}

Let $w_{i\ell}$ be supplier $i$'s contribution to 1995 unweighted mean width at release level $\ell$. The cumulative rank cost is $k_{i\ell}\in\{0,8,26,233\}$. Supplier independence gives the multiple-choice knapsack problem
\begin{equation}
\min_{\ell_1,\ldots,\ell_{14}}\sum_i w_{i\ell_i}
\quad\text{subject to}\quad\sum_i k_{i\ell_i}\leq B.
\end{equation}
Define $\Phi_0(B)=0$ for $B\geq0$ and $\Phi_i(B)=+\infty$ for $B<0$. The dynamic-programming recursion
\begin{equation}
\Phi_i(B)=\min_{\ell\in\{0,1,2,3\}}
\{w_{i\ell}+\Phi_{i-1}(B-k_{i\ell})\}
\end{equation}
is exact because an optimal final choice leaves the preceding suppliers optimal under the remaining budget.

Budget ceilings are rounded down to integers. Width-per-cost upgrades use 1995 gains, and purchase priority uses 1995 purchases. Random allocations sequentially add affordable upgrades (seed 20260815), and feasible allocations need not be equiprobable.

At the half-budget ceiling, purchase priority uses 1,631 statistics but leaves mean width at 14.19 points, whereas historical width minimization uses 1,606 and leaves 2.09 points (Table \ref{tab:network_measurement_full}). The difference persists when widths are weighted by purchases, so spending the budget on the largest flows need not measure the target exposures well.

\begin{table}[!htbp]
\centering\small
\caption{Information use and exposure uncertainty under alternative measurement rules}
\label{tab:network_measurement_full}
\begin{tabular}{@{}rlrrrrr@{}}
\toprule
\thead{Budget\\(\%)} & Rule & \thead{Additional\\statistics} & \thead{Mean\\width (pp)} & \thead{Weighted\\width (pp)} & \thead{Identified\\at 2\% (\%)} & \thead{Identified\\at 3\% (\%)} \\
\midrule
5 & 1995 optimal & 158 & 10.49 & 7.87 & 13.9 & 20.7 \\
5 & Width per cost & 148 & 11.01 & 8.38 & 13.0 & 20.7 \\
5 & Uniform & 112 & 12.89 & 9.56 & 11.1 & 16.8 \\
5 & Purchase priority & 156 & 21.39 & 14.74 & 2.9 & 4.3 \\
5 & Random median & 160 & 18.22 & 13.40 & 10.1 & 13.9 \\
\addlinespace[0.45em]
10 & 1995 optimal & 310 & 8.23 & 5.54 & 24.5 & 36.1 \\
10 & Uniform & 112 & 12.89 & 9.56 & 11.1 & 16.8 \\
10 & Purchase priority & 319 & 23.14 & 17.10 & 1.0 & 4.3 \\
10 & Random median & 323 & 18.81 & 14.26 & 11.1 & 14.2 \\
\addlinespace[0.45em]
25 & 1995 optimal & 778 & 5.45 & 3.80 & 38.9 & 51.0 \\
25 & Uniform & 364 & 7.87 & 5.12 & 29.3 & 38.9 \\
25 & Purchase priority & 811 & 15.49 & 9.83 & 5.8 & 5.8 \\
25 & Random median & 778 & 7.61 & 4.93 & 29.8 & 39.9 \\
\addlinespace[0.45em]
50 & 1995 optimal & 1,606 & 2.09 & 1.43 & 64.9 & 72.1 \\
50 & Uniform & 364 & 7.87 & 5.12 & 29.3 & 38.9 \\
50 & Purchase priority & 1,631 & 14.19 & 9.54 & 9.6 & 6.7 \\
50 & Random median & 1,606 & 4.57 & 2.98 & 42.8 & 51.9 \\
\addlinespace[0.45em]
100 & All rules & 3,262 & 0.00 & 0.00 & 100.0 & 100.0 \\
\bottomrule
\end{tabular}
\begin{minipage}{0.98\textwidth}
\vspace{0.4em}\footnotesize
\textit{Notes}: Additional statistics count template rows beyond baseline margins. Actual use may be below the budget ceiling. Unidentified dimensions in the full 26-supplier network equal 48,464 minus this count before reductions from accounting-forced zeros. Widths are in exposure percentage points, with weighted width using buyer purchases. The last two columns report the percentage of classifications identified at each threshold. All agree with the benchmark. Width-per-cost and 1995-optimal fields coincide at budgets of 10 percent and above. All rules coincide at full disclosure. Random entries are columnwise medians of 200 allocations and need not describe one allocation.
\end{minipage}
\end{table}

Holding the selected fields fixed, mean width at the quarter-budget ceiling rises from 4.84 points in 1995 to 4.87 in experimental 2000 and 5.45 in official 2005.

\subsection{Threshold conclusions and ranking reversals}
\label{app:measurement_certificates}

For a fixed buyer and threshold, the candidate measurements are the buyer's destination--group cell for each manufacturing supplier. Let $\ell_i^0$ be the baseline lower endpoint and $g_i\geq0$ its improvement from this cell. Order gains as $g_{(1)}\geq\cdots\geq g_{(14)}$. The minimum number of cells establishing an above-threshold conclusion is the smallest $k$ satisfying
\begin{equation}
\sum_i\ell_i^0+\sum_{h=1}^k g_{(h)}\geq\kappa.
\end{equation}
No selection of $k-1$ candidate cells can suffice, since its gain cannot exceed the sum of the largest $k-1$ gains. To establish a below-threshold conclusion, instead order the reductions in the upper bound. Additivity requires supplier-separable measurements.

\begin{samepage}
Duality makes the supporting accounting restrictions explicit. Consider the normalized program $\min_x c'x$ subject to $Ax=y$ and $0\leq x\leq d$. If $(\lambda,z)$ satisfies $A'\lambda+z\leq c$ and $z\leq0$, it gives the lower bound $\lambda'y+z'd$. At the optimum this equals the primal endpoint. Dual certificates for upper bounds apply the same construction to $-c$. The archived primal allocations and dual multipliers attain the endpoints. Sparse dual support need not be globally minimal.\par
\end{samepage}

To show that a selected buyer can fall outside the top twenty, we supply an allocation that matches all released totals and places twenty competitors above that buyer. The archived allocations cover every partial-budget selection under the deterministic rules and random allocations examined. Allocations for different buyers can use different networks and do not imply their simultaneous removal.

\subsection{Calculating maximum regret}
\label{app:monitoring_calculation}

Fix a selected set $S$ of size $K$ using the information released to the agency. Write $e_b=100E_b(f)/K$, and let $L_b$ and $U_b$ be valid marginal bounds in these units. Binary indicators $z_b$ select a competing set of size $K$, so $\sum_bz_b=K$. Introduce continuous variables $h_b=z_b(e_b-L_b)$. This product is imposed exactly by
\begin{align}
0&\leq h_b\leq (U_b-L_b)z_b,\\
h_b&\leq e_b-L_b,\\
h_b&\geq e_b-U_b+(U_b-L_b)z_b.
\end{align}
We calculate maximum regret using the mixed-integer linear program
\begin{equation}
\max_{f\in\Fcal(y),z,h}\left\{
\sum_b(L_bz_b+h_b)-\sum_{b\in S}e_b\right\}.
\label{eq:monitoring_milp}
\end{equation}
All buyers share the same allocation $f$. For fixed $f$, optimizing $z$ selects its top $K$ exposures, which proves equivalence to \eqref{eq:worst_monitoring_loss}. Compactness gives attainment. Binary shocked-origin flows can be extended to full origin allocations because the complementary origin margins form independent transportation tables within each released block.

For the coal score in \eqref{eq:coal_score}, replace $e_b$ by $(H E_b(f)-d_b)/K$ and use bounds in the same units. The score remains affine in $f$, so the product constraints and objective apply unchanged, with loss measured in days. Known Wyoming purchases bound Campbell flows, and complementary transportation tables separately allocate other Wyoming and non-Wyoming origins. The inventory-error sensitivity holds eligibility and the selected set fixed and permits independent changes in $d_b$. Its worst inventory vector raises coverage for selected plants and lowers it for unselected plants. A buyer common to both sets cancels from the loss.

We compare the global upper bound with an allocation refined by a continuous program that fixes the comparator set. Every reported gap for evaluating a specified set is below $10^{-5}$ in the relevant score units.

Marginal bounds give a conservative upper bound on maximum regret:
\begin{equation}
\max_{|T|=K}\left\{\sum_{b\in T\setminus S}U_b
-\sum_{b\in S\setminus T}L_b\right\}.
\end{equation}
It equals maximum regret only when the relevant marginal extrema can be attained together. Under the group release and the calibrated $\delta$, this bound is 4.429 points, compared with maximum regret of 4.132. At half the historical measurement budget, the corresponding values are 5.722 and 4.834.

\subsection{Choosing a monitoring set at a fixed release}
\label{app:monitoring_choice}

The half-budget comparison holds the 1,606 historically selected release statistics fixed and uses their 2005 values. It retains the 208 eligible buyers, capacity $K=20$, Tohoku manufacturing disturbance, and accounting restrictions of Table \ref{tab:monitoring_loss}. The objective is
\begin{equation}
R^*(y)=\min_{S:\,|S|=K}R(S;y).
\label{eq:minimax_monitoring}
\end{equation}
We choose a deterministic monitoring set and use withheld benchmark exposures only to evaluate it after selection.

We solve a sequence of finite-scenario problems. For feasible networks $f^1,\ldots,f^H$, binary indicators $s_b$ choose the monitored buyers, and the problem is
\begin{equation}
\begin{split}
\min_{s,\eta}\quad &\eta\\
\text{subject to}\quad &\sum_b s_b=K,\qquad s_b\in\{0,1\},\qquad \eta\geq0,\\
&\eta\geq\frac{100}{K}\left\{\max_{|T|=K}V(T,f^h)
-\sum_b s_bE_b(f^h,q)\right\},\quad h=1,\ldots,H.
\end{split}
\end{equation}
Restricting the comparison to these networks gives a lower bound on $R^*(y)$. Evaluating the chosen set over all networks allowed by the accounting restrictions gives an upper bound and a worst-case allocation to add to the next problem. We initialize with the proportional allocation and the worst-case allocation for its monitoring set, both computed from the releases. At each iteration, we retain the evaluated set with the lowest certified upper bound, including the proportional monitoring set among the candidates. Selection uses neither benchmark losses nor withheld exposures, including when breaking ties.

The search stops when the global gap is at most $10^{-5}$ points, or after 100 iterations or 60 minutes, whichever occurs first. Each mixed-integer solve is limited to 180 seconds and 20,000 nodes, subject to the remaining overall time. A finite-scenario solver bound remains a valid lower bound when a time or node limit prevents completion. Error-status bounds are discarded. An evaluated set supplies an upper bound even when minimax optimality is unresolved. Table \ref{tab:monitoring_choice} compares the proportional monitoring set with the best set found.

\begin{table}[!htbp]
\centering\small
\caption{Monitoring choices under the same half-budget release}
\label{tab:monitoring_choice}
\begin{tabular}{lrrr}
\toprule
Monitoring rule & \thead{Overlap with\\completion (\%)} & \thead{Benchmark\\loss (pp)} & \thead{Maximum\\regret (pp)} \\
\midrule
Proportional completion & 100 & 0.070 & 4.834 \\
Best set found & 50 & 0.783 & 3.285 \\
\bottomrule
\end{tabular}
\begin{minipage}{0.96\textwidth}
\vspace{0.4em}\footnotesize
\textit{Notes}: Both rules use the same 1,606 statistics, 208 eligible buyers, and capacity twenty. Losses are percentage points of mean exposure. Overlap is the percentage of the proportional rule's twenty buyers retained. Benchmark loss uses withheld exposures only after selection. The best set found has the lowest certified upper bound among the evaluated sets. The remaining bounds on the minimax value are reported below.\par
\end{minipage}
\end{table}

The search reaches the prespecified limit of 100 iterations. The bounds on $R^*(y)$, rounded outward, are [2.615, 3.286] points. The gap calculated before rounding is 0.669 points. This gap concerns the choice of monitoring set, whereas the numerical gap for evaluating each reported set is below $10^{-5}$ points. The saved worst-case allocations attain the unrounded upper bounds within that tolerance.

\section{Inference with Estimated Releases}
\label{app:inference}

\subsection{A polyhedral release region}

A polyhedral simultaneous confidence region for the release $y_0=Af_0$ can be written
\begin{equation}
\Ycal_n(1-\alpha)=\{y:D_ny\leq d_n\}.
\label{eq:release_region}
\end{equation}
For finite-sample validity, one can construct a valid interval for each released total and use a Bonferroni adjustment to ensure joint coverage. A less conservative asymptotic construction can use a max statistic based on survey replicate weights or influence functions. The procedure must reproduce the actual sampling design.

When the group shares in \eqref{eq:binary_restriction} are estimated, we include each common-allocation amount as a nuisance release in the joint confidence region. Absolute-value constraints remain linear for each fixed $\delta$.

\subsection{Binomial sampling experiment}
\label{app:binomial_sampling}

The binomial experiment in Section \ref{sec:use} generates 378 positive-purchase strata, of which 336 enter outside-Tohoku targets. Ten relevant strata have population share zero and none has share one. Their exact binomial intervals retain sampling uncertainty at the boundary.

Exact Clopper--Pearson intervals receive a target-specific Bonferroni adjustment. For each buyer, we count only suppliers whose inputs it purchases, and the count ranges from eight to fourteen.

Tables \ref{tab:assumption_precision} and \ref{tab:threshold_certification} give interval and classification results. Without sampling uncertainty, mean population width is 2.255 percentage points under the calibrated $\delta$ and 7.869 under accounting alone. The completion estimator's mean sampling variance falls from 0.0372 to 0.0004 squared percentage points between the smallest and largest samples, while its mean absolute population bias remains 0.174 percentage points.

\begin{table}[H]
\centering\small\setstretch{1.05}\renewcommand{\arraystretch}{1.05}
\caption{Interval width and coverage in the binomial experiment}
\label{tab:assumption_precision}
\begin{tabular}{@{}rlrrrr@{}}
\toprule
$n$ & Restriction & \thead{Mean width\\(pp)} & \thead{Coverage\\(\%)} & \thead{Weighted\\coverage (\%)} & \thead{Empirical joint\\coverage (\%)} \\
\midrule
500 & Common & 1.451 & 92.5 & 96.5 & 0.0 \\
500 & 1995 calibration & 3.944 & 100.0 & 100.0 & 100.0 \\
500 & Accounting & 12.916 & 100.0 & 100.0 & 100.0 \\
\addlinespace[0.45em]
2,000 & Common & 0.682 & 76.9 & 82.8 & 0.0 \\
2,000 & 1995 calibration & 3.076 & 100.0 & 100.0 & 100.0 \\
2,000 & Accounting & 10.193 & 100.0 & 100.0 & 100.0 \\
\addlinespace[0.45em]
10,000 & Common & 0.296 & 51.5 & 65.7 & 0.0 \\
10,000 & 1995 calibration & 2.606 & 100.0 & 100.0 & 100.0 \\
10,000 & Accounting & 8.837 & 100.0 & 100.0 & 100.0 \\
\addlinespace[0.45em]
50,000 & Common & 0.131 & 30.0 & 36.1 & 0.0 \\
50,000 & 1995 calibration & 2.402 & 100.0 & 100.0 & 100.0 \\
50,000 & Accounting & 8.278 & 100.0 & 100.0 & 100.0 \\
\bottomrule
\end{tabular}
\begin{minipage}{0.98\textwidth}
\vspace{0.4em}\footnotesize
\textit{Notes}: Each row uses 1,000 replications and 208 targets. $n$ is the binomial count per stratum. Widths are in exposure percentage points. Coverage concerns benchmark exposure, with weighted coverage using intermediate purchases. The 1995 calibration gives $\delta=0.0840847$. Empirical joint coverage is the percentage of replications covering all 208 targets. Intervals adjust over each target's active suppliers, so this frequency is not a nominal simultaneous confidence guarantee. Common-sourcing intervals exclude at least one benchmark target in every replication.
\end{minipage}
\end{table}

\begin{table}[H]
\centering\small\setstretch{1.05}\renewcommand{\arraystretch}{1.05}
\caption{Threshold classifications in the binomial experiment}
\label{tab:threshold_certification}
\begin{tabular}{@{}rlrrrr@{}}
\toprule
$n$ & Restriction & \thead{Classified\\at 2\% (\%)} & \thead{Incorrect\\at 2\% (\%)} & \thead{Classified\\at 3\% (\%)} & \thead{Incorrect\\at 3\% (\%)} \\
\midrule
500 & Common & 66.4 & 0.00 & 83.0 & 0.44 \\
500 & 1995 calibration & 25.5 & 0.00 & 45.4 & 0.00 \\
500 & Accounting & 13.1 & 0.00 & 22.0 & 0.00 \\
\addlinespace[0.45em]
2,000 & Common & 84.6 & 0.04 & 94.0 & 0.75 \\
2,000 & 1995 calibration & 38.9 & 0.00 & 58.3 & 0.00 \\
2,000 & Accounting & 19.9 & 0.00 & 29.6 & 0.00 \\
\addlinespace[0.45em]
10,000 & Common & 92.3 & 0.79 & 97.7 & 1.03 \\
10,000 & 1995 calibration & 48.3 & 0.00 & 64.4 & 0.00 \\
10,000 & Accounting & 23.9 & 0.00 & 34.0 & 0.00 \\
\addlinespace[0.45em]
50,000 & Common & 95.9 & 2.16 & 99.5 & 1.44 \\
50,000 & 1995 calibration & 52.6 & 0.00 & 66.7 & 0.00 \\
50,000 & Accounting & 26.8 & 0.00 & 36.0 & 0.00 \\
\bottomrule
\end{tabular}
\begin{minipage}{0.97\textwidth}
\vspace{0.4em}\footnotesize
\textit{Notes}: Entries are percentages of 208,000 target--replication pairs. A classification is reported when the interval lies entirely on one side of the threshold. An incorrect classification contradicts benchmark exposure. The 1995 calibration gives $\delta=0.0840847$.
\end{minipage}
\end{table}

\subsection{Multinomial releases and breakdown inference}

We next allow both origin and buyer margins to be estimated. Panel A of Table \ref{tab:inference} samples the Kanto transportation-equipment supplier block for goods buyers in the original 29-sector table. Write $\pi_{rj}=F_{ir,dj}/G$ for its $9\times18$ multinomial cell probabilities, where $G=9,359,510$ million yen. The unknown row and column probabilities are estimated jointly from the sample. We form exact Clopper--Pearson coordinate intervals with a Bonferroni allocation over these 27 margins. We then intersect this region with the adding-up restrictions. The plug-in procedure instead fixes them at the sample proportions.

For each goods buyer, the population margins imply an interval for Tohoku's contribution $\pi_{Tj}$ in this supplier block. Panel A counts a replication as covering only when all eighteen of these population intervals lie inside their respective projected intervals. This criterion concerns the separate scalar intervals, rather than the joint set of network links. Mean probability width averages over the eighteen coordinates and uses $G$ as its normalization.

For Kanto transportation equipment, total buyer purchases $D=16,097,139$ and the block scale $G$ are held fixed. Its input contribution in exposure percentage points is $100(G/D)\pi_{T,j^*}$. Establishing exposure above 2 percent requires $\pi_{T,j^*}>0.02D/G=0.034397$. The population interval for this contribution is $[2.265471,2.341814]$ percentage points. Table \ref{tab:inference} reports its projected width separately from the probability widths. At 100,000 draws, the mean projected width is 0.317126 points, compared with a population width of 0.076343 points. 

Panel B uses the Hokkaido electrical-machinery conclusion. Fourteen manufacturing supplier blocks are active in its sharp lower endpoint. Within each block, a binomial sample estimates the allocation between this buyer and the remaining buyers in its group. The experiment compares the plug-in breakdown estimate with the projected lower confidence bound in \eqref{eq:breakdown_lcb}.

\begin{table}[H]
\centering
\caption{Monte Carlo inference with estimated releases}
\label{tab:inference}
\small
\textit{Panel A. Multinomial releases and Kanto exposure}\par\vspace{0.4em}
\begin{tabular}{@{}rrrrrrr@{}}
\toprule
$n$ & \thead{Release\\coverage (\%)} & \thead{Plug-in\\coverage (\%)} & \thead{Projection\\coverage (\%)} & \thead{Mean cell width\\(probability)} & \thead{Buyer interval\\width (pp)} & \thead{Above threshold\\(\%)} \\
\midrule
1,000 & 98.3 & 0.0 & 99.9 & 0.009702 & 2.818 & 0.0 \\
5,000 & 98.2 & 0.0 & 99.9 & 0.002663 & 1.203 & 1.2 \\
25,000 & 97.6 & 0.0 & 99.8 & 0.000929 & 0.564 & 53.2 \\
100,000 & 97.5 & 0.0 & 99.7 & 0.000470 & 0.317 & 100.0 \\
\bottomrule
\end{tabular}
\par\vspace{0.9em}\textit{Panel B. Lower confidence bound for the breakdown point}\par\vspace{0.4em}
\begin{tabular}{@{}rrrrrr@{}}
\toprule
\thead{$n$ per\\stratum} & \thead{Release\\coverage (\%)} & \thead{Lower-bound\\coverage (\%)} & \thead{Mean plug-in\\$\widehat\delta$} & \thead{Mean lower\\bound} & \thead{Lower bound\\$\geq0.06$ (\%)} \\
\midrule
500 & 96.6 & 100.0 & 0.1371 & 0.0080 & 1.4 \\
2,000 & 94.9 & 99.8 & 0.1344 & 0.0452 & 23.5 \\
10,000 & 96.1 & 100.0 & 0.1329 & 0.0861 & 98.8 \\
50,000 & 95.9 & 100.0 & 0.1326 & 0.1108 & 100.0 \\
\bottomrule
\end{tabular}
\begin{minipage}{0.98\textwidth}
\vspace{0.4em}
\footnotesize
\textit{Notes}: Each row uses 1,000 replications. Panel A samples a multinomial flow table and forms Bonferroni simultaneous exact intervals for row and column releases. Projection coverage is simultaneous inclusion of the eighteen scalar Tohoku cell-probability intervals. Mean cell width is in probability units, and buyer interval width is in exposure percentage points for the named buyer. The above-threshold column reports the percentage of replications with the projected lower endpoint above the threshold. Panel B samples supplier blocks separately and constructs a one-sided projected lower confidence bound for the breakdown point. Nominal release confidence is 95 percent.
\end{minipage}
\end{table}

Projection covers all eighteen scalar population intervals in at least 99.7 percent of the reported multinomial replications, while the plug-in procedure almost never includes them all. Bonferroni projection is conservative at small samples, and the probability that the interval lies above the threshold rises as the release becomes more precise.

In Panel B, the mean lower confidence bound on the breakdown point rises from 0.0080 to 0.1108 as the sample per stratum increases from 500 to 50,000. The probability that this lower confidence bound is at least $0.06$ rises from 1.4 to 100 percent.

\section{Domestic Propagation and Sufficient Aggregate Information}
\label{app:network_propagation}

\subsection{Import allocation and gross outputs}

To form domestic propagation coefficients, we remove foreign imports from the published table's own-region cells. In the archived 53-sector workbook, $U_{rk}$ denotes own-region use for detailed supplier $k$ in region $r$ (column code 630). Let $M_{rk}$ denote the positive import deduction, equal to the negative of column code 730. The latter equals imports plus customs duties and import commodity taxes, column codes 700, 710, and 720. The observed total $U_{rk}$ equals intermediate use plus final use, codes 540 and 620. These identities hold for every retained detailed region--supplier pair.

The allocation assumption is a common foreign-import share across own-region users of a detailed input:
\begin{equation}
\mu_{rk}=M_{rk}/U_{rk},\qquad
\widetilde M_{rk,r\ell}=\mu_{rk}F^{\rm raw}_{rk,r\ell}.
\end{equation}
All observed rates lie in $[0,1]$. This proportional treatment follows METI's construction report \citep{METIConstruction2010}. We apply it to signed detailed uses and aggregate with the common-sector crosswalk. Let $\widetilde M_{idj}$ denote the resulting import allocation.

After aggregation, let $F^{\rm tab}$ denote the nonnegative table used for direct exposure. Balancing entries can make an allocated import amount negative or exceed the corresponding cleaned cell. We reconcile these entries by
\begin{equation}
M_{idj}=\min\{F^{\rm tab}_{i\,d\,d\,j},
\max(0,\widetilde M_{idj})\},\qquad
F^D_{irdj}=F^{\rm tab}_{irdj}-\mathbf1\{r=d\}M_{idj}.
\end{equation}
The lower-bound adjustment adds 23.061 million yen to allocated imports, while the upper-bound adjustment removes 1,151.529 million yen. Together their absolute magnitude is below 0.003 percent of allocated intermediate imports. Domestic and imported inputs are nonnegative and sum cell by cell to the table used for direct exposure.

Retained intermediate purchases total 455,065,764 million yen. Allocated imported inputs are \ImportedInputTotal\ million yen, and domestic inputs are \DomesticInputTotal\ million yen. Omitted activities have zero disturbances in the propagation experiment. Interregional flows are unchanged by the allocation. The endogenous domestic block retains Others.

Gross outputs $X$ come from column sector code 760 and destination-total code 10, used once per source region--industry. Each of the 234 aggregated outputs equals both the sum over destination codes 1--9 and gross inputs. We preserve these observed outputs when normalizing domestic purchases. The residual of gross output minus retained inputs includes accounting and omitted categories. It is not separately identified value added.

\subsection{Conditional next-round identification}

The baseline fixes each buyer's domestic purchases by supplier industry and national supplier-origin totals. The full-group release adds origin--destination--buyer-group flows for all 26 suppliers. It fixes $\sum_{b\in g}X_bv_b$ in each of the 27 groups, explaining the rank count in Section \ref{sec:propagation}. A support calculation verifies this increment for all nine manufacturing shocks.

For the outer bounds in Table \ref{tab:network_second}, we propagate valid coordinate envelopes. Conditional on measured $v$, each sharp endpoint is attained by a domestic table preserving all group flows and direct responses.

\begin{figure}[!htbp]
\centering
\includegraphics[width=0.98\textwidth]{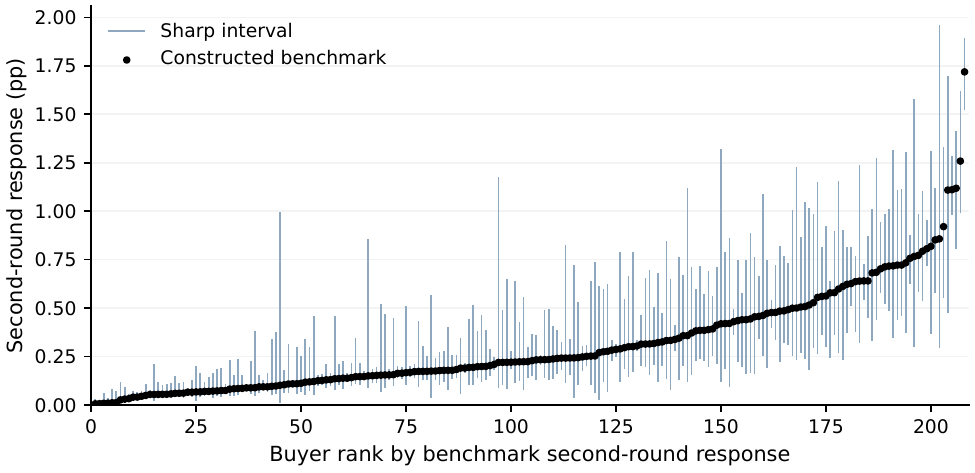}
\caption{Second-round intervals after measuring all direct responses}
\label{fig:network_propagation}
\begin{minipage}{0.95\textwidth}
\vspace{0.3em}\footnotesize
\textit{Notes}: Vertical lines give sharp intervals for the 208 outside-Tohoku buyers, ordered by their benchmark second-round response, with dots marking the benchmark. Each endpoint is attained by a compatible domestic network, which can differ across buyers and endpoints. Responses are percentage points per unit Tohoku manufacturing cost disturbance. The release contains all domestic buyer-group totals and every direct response. Foreign input prices are fixed.
\end{minipage}
\end{figure}

Second-round uncertainty remains throughout the distribution of benchmark responses (Figure \ref{fig:network_propagation}). Across shock origins, mean width ranges from 0.034 points for Okinawa to 1.455 for Kanto (Table \ref{tab:domestic_regional_second}).

\begin{table}[!htbp]
\centering\small
\caption{Sharp domestic second-round widths by shock origin}
\label{tab:domestic_regional_second}
\begin{tabular}{@{}lr@{}}
\toprule
Shock origin & Mean sharp width (pp) \\
\midrule
Hokkaido & 0.254 \\
Tohoku & 0.360 \\
Kanto & 1.455 \\
Chubu & 0.769 \\
Kinki & 1.005 \\
Chugoku & 1.081 \\
Shikoku & 0.347 \\
Kyushu & 0.692 \\
Okinawa & 0.034 \\
\bottomrule
\end{tabular}
\begin{minipage}{0.91\textwidth}
\vspace{0.4em}\footnotesize
\textit{Notes}: Mean sharp second-round widths across 208 outside-origin buyers, in percentage points per unit manufacturing cost disturbance. Every direct response and all domestic origin--destination--buyer-group totals are observed. Domestic inputs use the proportional import allocation described above, with foreign input prices held fixed.
\end{minipage}
\end{table}

\subsection{Sufficient aggregate information for several rounds}
\label{app:invariant_span}

Observations $v_0=q,v_1,\ldots,v_K$ satisfying $\Omega v_{h-1}=v_h$ identify the first $K$ rounds. What information would identify all subsequent rounds? It is sufficient to identify how the network transforms a space of shocks containing the initial disturbance, provided each transformation remains within that space. The following condition makes this requirement precise.

Let $Q$ and $H$ be known matrices, with $Q$ of full column rank. Suppose the measurements imply $\Omega Q=QH$ for every feasible network. The columns of $Q$ then span a space of shocks on which the network's action is known. For $q=Qa$, induction gives $\Omega^kq=QH^ka$. Stability then yields $u=Q(I-H)^{-1}a$. This invariant-subspace condition is related to lumpability for partition matrices \citep{TianKannan2006}.

Pooled group flows generally reveal group averages, which need not identify $\Omega Q$ for individual buyers. Related work uses a rank-one approximation to study upstreamness and downstreamness from local and aggregate accounts \citep{BartolucciEtAl2024}. The condition above gives exact identification across the accounting set.

\subsection{Uniform stability from domestic purchases}

Let $w_i>0$ be industry weights, extended to nodes by industry, and let $c^D_{ib}$ be observed domestic purchases. If, for every buyer $b$,
\begin{equation}
\sum_i\frac{c^D_{ib}}{X_b}w_i\leq\lambda w_{j(b)},\qquad \lambda<1,
\label{eq:uniform_network_stability}
\end{equation}
then every feasible nonnegative network satisfies $\Omega w\leq\lambda w$. Its weighted supremum norm and spectral radius are bounded by $\lambda$ uniformly over the domestic accounting set.

Positive weights satisfy all 234 inequalities at $\lambda=\DomesticContraction$. Unweighted sums exceed one only for the nine Others rows (maximum \DomesticMaximumRow) and do not establish stability. The benchmark spectral radius \DomesticSpectralRadius\ concerns that network alone.

\end{document}